\documentclass[amsmath,amssymb,twocolumn,aps,prb,superscriptaddress]{revtex4-2}
\usepackage[utf8]{inputenc}
\usepackage[normalem]{ulem}
\usepackage{physics}
\usepackage{epsfig,subfigure,amsmath,float,xcolor,bm}
\usepackage[colorlinks]{hyperref}
\hypersetup{colorlinks=true,linkcolor=blue,
filecolor=blue,urlcolor=blue,citecolor=blue}
\definecolor{hgt}{rgb}{0.,0.4,0.4}
\newcommand{\sig}{\si}
\newcommand\beq{\begin{equation}}
\newcommand\eeq{\end{equation}}
\newcommand\bes{\begin{subequations}}
\newcommand\ees{\end{subequations}}
\newcommand\bea{\begin{eqnarray}}
\newcommand\eea{\end{eqnarray}}
\newcommand\non{\nonumber}

\newcommand\ig{\includegraphics}

\newcommand\al{\alpha}
\newcommand\be{\beta}

\newcommand\Ga{\Gamma}

\newcommand\De{\Delta}

\newcommand\si{\sigma}

\newcommand\lam{\lambda}

\newcommand\ua{\uparrow}
\newcommand\da{\downarrow}

\newcommand{\Va}{{\bf a}}
\newcommand{\Vr}{{\bf r}}
\newcommand\vk{{\bf k}}
\newcommand\til{\tilde}
\newcommand{\mH}{\mathcal{H}}

\newcommand{\RNum}[1]{\uppercase\expandafter{\romannumeral #1\relax}}

\begin{document}
\title{Josephson Junction of Nodal Superconductors with Rashba and Ising Spin-Orbit coupling}
\author{Gal Cohen}
\email{galao@post.bgu.ac.il}
\affiliation{Department of Physics, Ben-Gurion University of the Negev,
Beer-Sheva 84105, Israel}
\author{Ranjani Seshadri}
\email{ranjanis@post.bgu.ac.il}
\affiliation{Department of Physics, Ben-Gurion University of the Negev,
Beer-Sheva 84105, Israel}
\author{Maxim Khodas}
\email{maxim.khodas@mail.huji.ac.il}
\affiliation{Racah Institute of Physics, Hebrew University of Jerusalem,
Jerusalem 91904, Israel}
\author{Dganit Meidan}
\email{dganit@bgu.ac.il}
\affiliation{Department of Physics, Ben-Gurion University of the Negev,
Beer-Sheva 84105, Israel}
\affiliation{Université Paris-Saclay, CNRS, Laboratoire de Physique des Solides, 91405, Orsay, France.}
\date{\today}
\begin{abstract}
We study the effect of a Rashba spin-orbit coupling on the nodal superconducting phase of an Ising
superconductor. Such nodal phase was 
predicted to occur when applying an  in-plane field beyond the Pauli limit to a superconducting monolayer transition metal dichalcogenides (TMD).
Generically, Rashba spin-orbit is known to lift the chiral symmetry that protects the nodal points, resulting in a fully gapped phase. However, when the magnetic field is applied along the $\Gamma -K $ line, a residual vertical mirror symmetry protects a nodal crystalline phase. We
study  a single-band tight-binding model that captures the low energy physics around the $\Gamma $ pocket of monolayer TMD. We calculate the topological properties, the edge state structure, and the current phase relation in a Josephson junction geometry of the nodal crystalline phase. We show that while the nodal crystalline phase is characterized  by localized edge modes on non-self-reflecting boundaries, 
the current phase relation exhibits a trivial $2\pi $ periodicity in the presence of Rashba spin-orbit coupling.

\end{abstract}
\maketitle

\section{Introduction}
Transition metal dichalcogenides (TMDs) such as ${\rm NbSe_2}$ and  ${\rm MoS_2}$ have been
proposed and experimentally confirmed  to be an ideal platform for in-depth explorations for unconventional superconductivity - 
both intrinsic and externally induced \cite{Zhou2016,Ilic2017,He2018,Sosenko2017,Nakamura2017,Mockli2018,Fischer2018,Smidman2017,Yuan2014,Oiwa2018,Hsu2017,Wang2018,Oiwa2019,Sohn2018}. 

More recently, cutting-edge advances in fabrication techniques have
facilitated the engineering of layered systems from these TMDs where the constituent layers are held
together by weak Van der Waals force\cite{Wang2012,Geim2013}. Here, some  systems are found to retain their 
superconducting property even down to the monolayer limit
\cite{Lu2015,Ugeda2015,Saito2016,Xi2016,Costanzo2016,Dvir2017,DelaBarrera2018,Sohn2018,Xing2017,Hamill2021}.

Unlike their bulk counterparts, many monolayer and few-layer TMD's break
inversion symmetry, thereby giving rise to a very strong Ising spin-orbit coupling (SOC) \cite{Wang2012,Yuan2014,Lu2015,
Ugeda2015,Xi2016,Costanzo2016, Saito2016, yuan2016ising,Sohn2018} which pins
the electron spins perpendicular to the plane. The most remarkable consequence of this strong SOC
is that
superconductivity survives at high in-plane magnetic fields even beyond the Pauli critical limit 
\cite{Lu2015,Saito2016,Ilic2017,Xi2016,Dvir2017,DelaBarrera2018,Sohn2018,Liu2018,Cho2020,cho22,Frigeri2006,Kuzmanovic2022}. 


It was proposed that the presence of an in-plane field can induce a topological transition into a nodal superconducting phase \cite{He2018,Fischer2018} protected by a combination of an effective time reversal  and particle-hole symmetry. The nodal superconducting phase is expected to be accompanied  by Majorana flat bands \cite{Zhao2013,Matsuura2013}, indication of which have been reported \cite{Galvis2014,Nayak2021}, as well as  distinct $4\pi $ periodic Josephson current for the transverse momenta in-between the nodal points \cite{sesh22}.


In this paper we study the effect  of Rashba SOC on  the nodal superconducting phase, focusing on the boundary modes and  the  Josephson current phase relation. Rashba SOC is naturally present due to electronic gates and the presence of a substrate and can be tuned experimentally. The presence of Rashba spin-orbit breaks the chiral symmetry that protects the nodal superconducting phase, and as a result, the nodal points  are generally gapped. However, when the in-plane field is aligned along the $ \Gamma -K$ direction, a lower crystalline symmetry protects the nodal phase \cite{fernandes20}. We study the boundary states in the crystalline phase as well as the current-phase relation in a Josephson-junction 
geometry. Our results indicate that  while the vertical mirror symmetry protects exponentially localized states at the boundary transformed by the symmetry, the current phase relation exhibits a trivial $2\pi $ periodicity in the presence of Rashba spin-orbit.

The plan of the paper is as follows. We begin in Sec. \ref{sec:continuum} with an analysis of the
low energy momentum-space Hamiltonian and its related  symmetries. In  Sec. \ref{sec:lattice} 
we  introduce a toy model on a triangular lattice which reduces to the
continuum Hamiltonian close to the $\Gamma-$ point. We discuss the topological properties of this model with and without Rashba spin-orbit and study the stability of the boundary modes in a ribbon 
geometry.
The physics of a Josephson junction fabricated out of such a material is discussed in Sec. \ref{sec:JJ}.

\section{Continuum Model } \label{sec:continuum}
An Ising superconductor such as monolayer NbSe$_2$  subjected to an in-plane
magnetic field of magnitude $h$, with a superconducting pairing $\Delta$ is governed by the following
Bogoliubov-de-Gennes (BdG) Hamiltonian,

\bea
\cal{H}(\vk)&=&\xi(\vk)\tau^z + \lam(\vk) \sig^z - \al_R(k_x\tau^z\sig^y-k_y\sig^x) \non \\
&~& + h \cos \theta \tau^z  \sig^x + h \sin\theta \tau^0 \sig^y  \non \\
&~& + \Re(\Delta) \tau^y \sig^y+ \Im(\Delta) \tau^x \sig^y, \label{eq:Hkk}
\eea

where $\xi(\vk) = (k_x^2 + k_y^2)/2m - \mu$ is the kinetic energy term with $\mu$ being the chemical
potential. The Ising SOC $\lam(\vk) = \lam_I(k_x^3 - 3k_x k_y^2)$ is unique to
this class of materials, and pins the electron spins perpendicular to the $x-y $ plane.  The form of $\lambda(\vk)$ is constrained by the crystalline symmetry point group $D_{3h}$ which includes 
a mirror reflection plane $M_z$ (with normal along the $z-$direction),
a three-fold rotational symmetry $ C_3$ and a vertical mirror $M_x$ (with normal along the $x-$direction).
The strong Ising SOC protects superconductivity in 
the presence of an in-plane magnetic field $h$ which can exceed the  Pauli limit. The parameter
$\theta$ denotes the angle the in-plane magnetic field makes with the $x-$axis.
$\al_R$ determines the strength of Rashba SOC, typically present in experimental setups,
and can be tuned by gating or by appropriate choice of substrate.

In the absence of Rashba SOC i.e. when $\al_R=0$, the in-plane direction of $h$ is immaterial. When
$|h|>\De$ the BdG spectrum has twelve nodal points on the high symmetry $\Ga-M$ lines
$k_x = 0, \pm\sqrt{3}$ along which the Ising SOC vanishes. This nodal superconducting phase is accompanied by the 
presence of Majorana flat bands \cite{He2018,Fischer2018,Zhao2013,Matsuura2013}, as well as an energy phase relation 
that depends on the momentum transverse to the current direction, with a $4\pi$ periodicity for
the momenta lying between each pair of nodal points \cite{sesh22}.

In this work we  analyze the effect of Rashba SOC on the nodal superconducting phase, the fate of its boundary modes, 
and the Josephson current phase relation. To this end, we work in a parameter regime where
$h>|\De|$ and there are twelve nodal points in the absence of Rashba SOC. 


\subsection {Family of 1D Hamiltonians and symmetries}
The origin of the nodal points can be understood by analyzing the family of $1D $ Hamiltonian
obtained by setting $k_y $ as a  parameter, $\mH^{(1D)}_{k_y}(k_x)$.
In the absence of Rashba SOC, i.e. when $\al_R=0$, this model has a particle-hole symmetry given by
\beq
C\mH^{(1D)}_{k_y}(k_x)C^{-1}=-\mH^{(1D)}_{k_y}(-k_x) \label{eq:phsymm}
\eeq
with $C = \tau^x K $. While the magnetic field explicitly breaks time-reversal symmetry,
the model has an emergent modified time-reversal (TR) symmetry,
\beq
T\mH^{(1D )}_{k_y}(k_x)T^{-1}=\mH^{(1D)}_{k_y}(-k_x)
\eeq
with $T = \sig^x\tau^z K= \Theta M_z\tau_z $ which is a combination of time-reversal symmetry
$\Theta = i\sigma_y K $ and  basal plane mirror symmetry $M_z$. The family of $1D$
Hamiltonians therefore lies in class BDI
of the Altland-Zirnbauer classification \cite{alt97}. The presence of the nodal points can
therefore be understood as a series of topological phase transitions tuned by the parameter
$ k_y$ as explained in Ref. \onlinecite{sesh22}. 
Next, we introduce a Rashba SOC as given in Eq. \eqref{eq:Hkk} which consists of two parts.
The first term $k_x\sig^y\tau^z $ breaks the modified time-reversal symmetry $T$ while the second
term $k_y\sig^x $ breaks particle hole symmetry $C$ of the effective $1D $ model, leaving $\mH^{1D}_{k_y}$ in class A.
However, when the field is oriented along the $x$-axis, i.e. when $\theta =0$, the system has a residual vertical mirror symmetry plane,  defined by,
\beq \label{eq_Mirror}
{M}_x \mH^{(1D)}_{k_y}(k_x){M}^{-1}_x=\mH^{(1D)}_{k_y}(-k_x)
\eeq

with ${M}_x = \si_x \tau_z$. We will show below that this 1D Hamiltonian realizes a crystalline
 insulating phase associated with gapless edge states which are localized along the $x-$direction and
propagate along $y-$direction. 

\section{Lattice Model}\label{sec:lattice}
To gain further insight into the topological phase and the  nature of its boundary modes, we study a tight-binding
model presented in \cite{Haim2022,smidman2017superconductivity}
that captures the key features of the topological superconducting phase and exhibits the same
low energy physics in the continuum limit $ka\rightarrow 0$.

\begin{figure}[htb]
\centering
\ig[width=.45\textwidth]{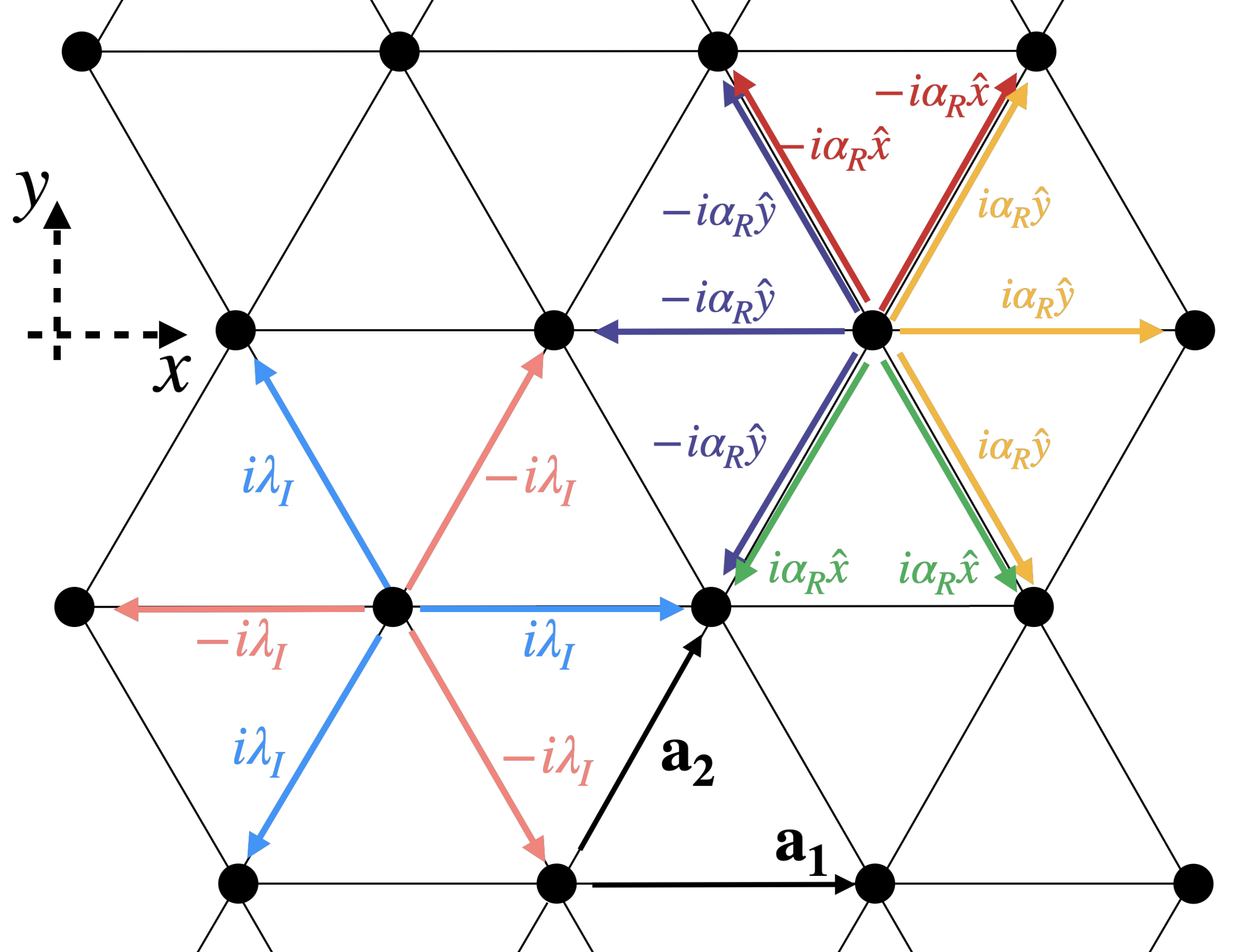}
\caption{Schematic diagram of the triangular lattice used for the tight-binding model showing the lattice
vectors ${\bf{a_{1,2}}}$ and the hopping amplitudes corresponding to the Ising and Rashba SOC. This hopping
profile results in the Hamiltonian $H_I$ and $H_R$ in Eq.\eqref{eq:HI} and Eq. \eqref{eq:HR} respectively.}

\label{fig:trilattice}
\end{figure}
The lattice model consists of a nearest neighbor hopping:
\bea \label{eq:H0}
H_0 = -t \sum_{<i,j>,s} {c^\dag_{i,s}} c_{j,s} -\tilde{\mu} \sum_{i,s}  {c^\dag_{i,s}} c_{i,s} 
\eea
where $s=\ua, \da$ denotes the spin, $<i,j>$ spans all the nearest neighbors
and $\tilde{\mu}$ is the on-site chemical potential. 
In the continuum limit, $ka\rightarrow 0 $, this reduces to the kinetic energy term $\xi(\vk)$ of Eq. \eqref{eq:Hkk} when we set
$\tilde{\mu} = \mu-6t$ and $t = 1/12m$. 
Similarly, Ising SOC is modeled as a nearest-neighbor hopping with alternating signs (shown in Fig.
\ref{fig:trilattice}), that reflect the $C_3$ symmetry. Note that the sign is opposite for the two spins.
\bea 
H_I = \frac{i \lam_{\rm I} }{2}\sum_{<i,j>,s,s'}\nu_{ij}\si^z_{ss'} {c^\dag_{i,s}} c_{j,s'}
\label{eq:HI}
\eea
where $\nu_{ij}=+1(-1)$ for ${\Vr}_{ij}= \Vr_i-\Vr_j = {\Va}_1, -{\Va}_2,{\Va}_2-{\Va}_1(-{\Va}_1,{\Va}_2,-{\Va}_2+{\Va}_1)$,
respectively, and the lattice vectors are: ${\Va}_1 = (2a,0) $ and ${\Va}_2 = {a}(1,\sqrt{3}) $. 

The in-plane magnetic field $({\bf h} = h_x,h_y)$ 
\beq\label{eq:HB}
H_B = \sum_{i,s,s'} \left(\bf{h}\cdot \bm{\si}\right)_{ss'}{c^\dag_{i,s}} {c_{i,s'}}
\eeq
Lastly, the Rashba term can be written as follows
\beq
H_R =  -\frac{i\al_R}{6}\sum_{<i,j>,s,s'} {\bf z}\cdot \left(\Vr_{ij}\times
{\bm\si} \right)_{ss'}{c^\dag_{i,s}} {c_{j,s'}} \label{eq:HR}
\eeq
In momentum space the lattice Hamiltonian, eq. \eqref{eq:H0}-\eqref{eq:HR}  take the following form
\bea\label{eq:LatH}
H&=&\sum_{\vk,s}\til{\xi}(\vk)c_{\vk}^\dag c_{\vk}-\sum_{\vk,ss'}\til\lam_I(\vk)\cdot{\bm\si}_{ss'}c_{\vk s}^\dag c_{\vk s'} \non \\
&+& \sum_{\vk,ss'}\til\al_R(\vk)\cdot{\bm\si}_{ss'}c_{\vk s}^\dag c_{\vk s'} + \sum_{\vk,ss'}{\bf h }\cdot{\bm\si}_{ss'}
c_{\vk ,s}^\dag c_{\vk ,s'}, \non \\
\eea
where the kinetic energy term is 
\beq
\til\xi(\vk) = -4t\cos(k_x)\cos(\sqrt{3}k_y)-2t\cos(2k_x)-\til\mu.
\eeq
 Here $\til\lam_I(\vk)$ and  $\til\alpha_R(\vk)$ correspond to the Ising and Rasbha SOCs respectively and have the following 
form in the lattice model,
\bea
\til\lam_I(\vk) = \lam_I \hat{z} [\sin(\vk \cdot {\Va}_1) + \sin(\vk \cdot ({\Va}_2-{\Va_1}))-\sin(\vk \cdot {\Va}_2)], \non \\
\eea
and
\bea
\til\al_R(\vk) & =&-\frac{\sqrt{3}\al_R}{2}\hat{x}[\sin(\vk \cdot ({\Va}_2-{\Va_1}))+\sin(\vk \cdot {\Va}_2)] \non\\
&-& \frac{\al_R}{2}\hat{y}[\sin(\vk \cdot ({\Va}_2-{\Va_1}))-\sin(\vk \cdot {\Va}_2)-2 \sin(\vk \cdot {\Va}_1)] \non \\
\eea
The strength of each term is suitably chosen to give the same low energy Hamiltonian as Eq. \eqref{eq:Hkk}.

\begin{figure}[htb]
\centering
\ig[width=.45 \textwidth]{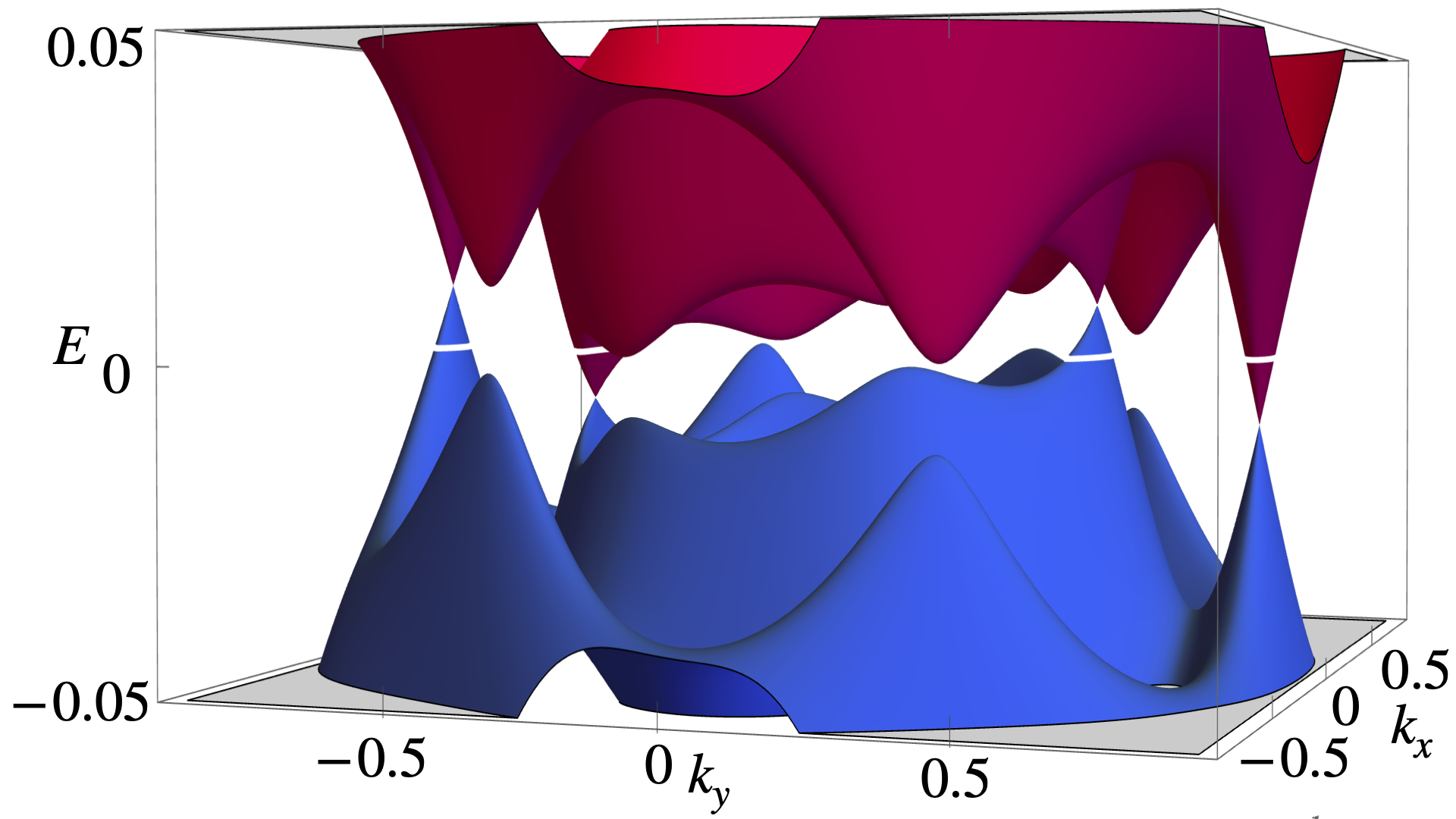}
\caption{Energy dispersion of the two low  energy bands of eq. \eqref{eq:LatH} in the vicinity of the $ \Gamma$ point.
The parameters used are  
$m=1$, $\tilde{\mu} = -0.3$, $\lam_{SO} =0.15$, $h=0.1 $, $ \Delta = 0.06$ and $ \al_R= 0.02 $. The
white line marks the zero-energy contour. The four gap-closing points are shifted away from the
$E=0$ plane and lie on the $k_x=0$ line.}
\label{fig:3DDisp1}
\end{figure}

Figure \ref{fig:3DDisp1} shows the dispersion of the two lower energy bands of the lattice model given by Eq. \eqref{eq:LatH},
in the vicinity of the $\Gamma $ point. The Rashba SOC breaks the chiral symmetry  thus generically lifting
the nodal points resulting in a fully gapped phase. However, 
when the magnetic field is aligned along the $\Gamma -K $ direction, the system has a residual vertical mirror symmetry $M_x $, 
which protects the nodal points along the $ k_x =0 $ symmetry line. The system therefore realizes a
nodal crystalline phase, as we show below. Due to the breaking of chiral symmetry, the nodes are shifted away from zero energy.   

\begin{figure}[htb]
\centering
\ig[width=.43\textwidth]{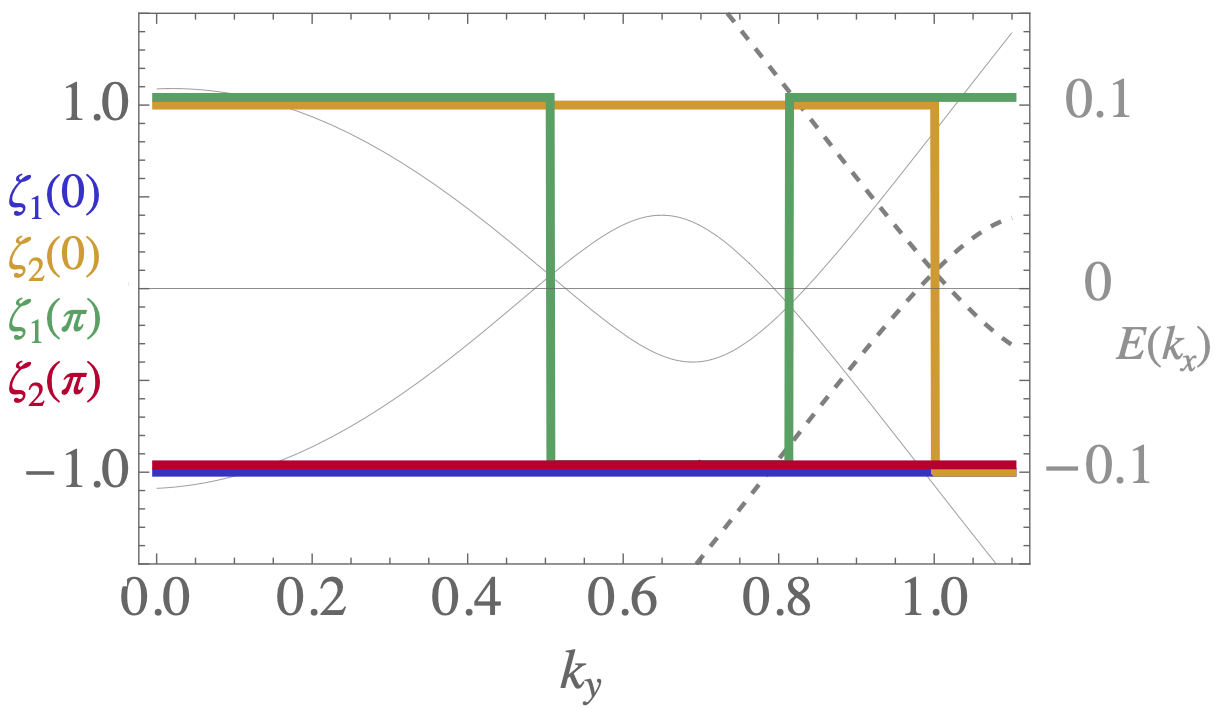}
\caption{The evolution of the Mirror eigenvalues at the reflection symmetric momenta $k_x=0,\pi $
of the two lowest energy levels for the lattice toy model (The lines are shifted vertically for clarity).
The dispersion of the two low energy bands at $k_x=0 $ ($k_x=\pi $) is shown in solid (Dashed) gray lines,
indicating the location of the nodal points. The parameters used are  
$m=1$, $\tilde{\mu} = -0.3$, $\lam_{SO} =0.15$, $h=0.1 $, $ \Delta = 0.06$ and $ \al_R= 0.02 $.  }
\label{fig:inversion}
\end{figure}

\subsection{Symmetries and topological classification}
To gain further insight into the topological aspects of the lattice model and the origin of the
nodal points in the spectrum we consider the family of lattice $1D$ Hamiltonians obtained by
treating $k_y $ in eq. \eqref{eq:LatH} as a parameter. 

As discussed in Sec. \ref{sec:continuum}, Rashba SOC lifts the chiral symmetry thus
leaving the family of $ 1D$  lattice  Hamiltonians  in class A. However, when the magnetic field is
aligned along the $ \Gamma - K$ the resulting $1D$ hamiltonian  is symmetric under vertical mirror,
${M}_x$, and is gapped except for 4 discrete  values of $k_y$.
For values of $ k_y$ between these nodal
points the system  realizes a one-dimensional topological crystalline phase
\cite{Hughes2011,Chiu2013}. 

At $k_x^{(\rm inv)} = 0,\pi $, the $1D $ Hamiltonian is mapped onto itself under reflection. In these
reflection symmetric momenta,  the energy levels have a well-defined reflection eigenvalue. The reflection
eigenvalues of the two occupied levels, labeled as  $ 1,2$ are  $\zeta_{1,2}(k_x=0) $ and $\zeta_{1,2}(k_x=\pi)$
corresponding to  $\Pi_{k_x=0}{M}_x\Pi_{k_x=0} $ and $\Pi_{k_x=\pi}{M}_x\Pi_{k_x=\pi} $, respectively,
where $ \Pi_{k}$ is the projector onto the two lowest energy levels
at a given momentum $k$. These are continuously connected to the negative energy states in the absence of Rashba SOC.
The reflection eigenvalues  define a $ \mathbb{Z}_2$ index given by:
\begin{eqnarray}
    \nu_{{\cal M}} = \prod_{i\in {\rm occ}, k_x^{\rm inv}} \zeta_{i}(k_x^{\rm inv}).
\end{eqnarray}

Figure \ref{fig:inversion} shows the value of the reflection eigenvalues of the two occupied bands $1,2 $ at
the two reflection symmetric momenta, $k_x^{(\rm inv)} = 0,\pi $, as a function of parameter $k_y$. Solid and
dashed gray lines indicate the  spectra along the $k_x=0$ and $k_x=\pi$ lines, respectively. The closing and
reopening of the band gap is accompanied by a topological phase transition; i.e. a change in sign of
$\zeta_1(\pi)$ (for the gap closing at $k_x =0 $) and $\zeta_2(0)$ (for the gap closing at $k_x =\pi $).

\subsection{Bulk boundary correspondence}
To examine the bulk boundary correspondence for the nodal crystalline phase, we 
study the  tight-binding lattice Hamiltonian on a ribbon-like geometry with open boundary conditions in
the (non self reflecting) $x-$direction, and periodic boundary conditions in the $y-$direction.
 This makes $k_y$ a good
quantum number and allows us to write the Hamiltonian in the ribbon geometry as an  effective $1D$  chain for a given $k_y $:
\beq
H_{\rm Rib}(k_y)=H_{0}(k_y)+H_{I}(k_y)+H_{R }(k_y)+H_{B}+H_{SC}.\label{eq:Hribbon}
\eeq
These terms correspond to kinetic energy, Ising SOC, Rashba SOC, in-plane field and
superconductivity respectively, a detailed expression is given in Appendix \ref{sec:ribbonH}.

\begin{figure}[htb]
\centering
\ig[width=.4\textwidth]{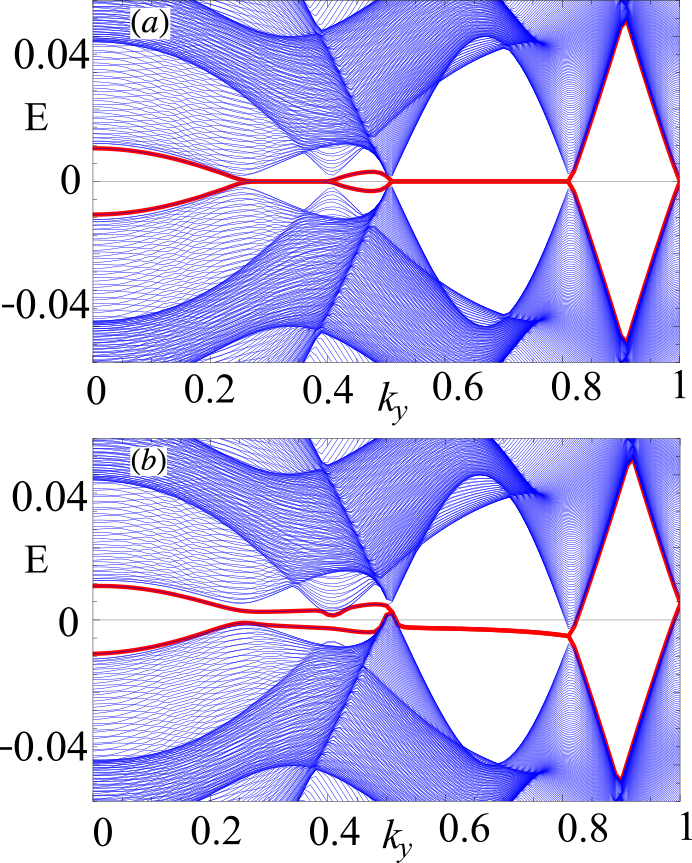}
\caption{Energy spectrum for a nano-ribbon with open boundary conditions
for (a) $\al_R=0$ and (b) $\al_R=0.01$ obtained by diagonalizing Eq. \eqref{eq:Hribbon}.
The Rashba SOC breaks Chiral symmetry
and moves the nodal points away from zero energy. Midgap states localized at the open boundaries  of the ribbon are
marked in red. These states decay exponentially into the bulk for momenta $k_y$ between pairs of nodal points.}
\label{fig:edgespec}
\end{figure}

Diagonalizing the Hamiltonian in Eq. \eqref{eq:Hribbon}, we obtain the eigenvalues and the corresponding
wavefunctions for the bulk and edge states. Figure  \ref{fig:edgespec} (a) shows the BdG spectrum in the
Ribbon geometry for $\al_R = 0$. 
Eigenvalues corresponding to states localized on the open $x-$direction boundary are shown in red. 
The two pairs of nodal points at $k_y\approx 0.25,0.4  $  and $k_y\approx 0.5,0.8 $ 
are accompanied by the appearance of zero energy states which are localized on the open $x-$direction boundary
(marked  in red).

Figure \ref{fig:edgespec}(b) shows the BdG spectra for $\al_R=0.01$.  When the Rashba
term is switched on, only one pair of  nodal points survive, which are   shifted away from zero energy. 
In Fig. \ref{fig:edgespec}(b) these are located at $k_y \approx 0.5$ and $0.8$. The two degenerate midgap
states connecting this pair of nodes (shown in red)  live on the (non-self-reflecting) x-boundary 
and have a non-zero  dispersion as a function of $k_y$ (the conjugate momenta for the direction parallel to the boundary). The degeneracy between the
boundary mode is protected by the mirror symmetry $M_x $.
In Appendix \ref{sec:invbreak} we show that any perturbation breaking this reflection symmetry splits 
these edge states, see Fig.~\ref{fig:inversion}. 

\begin{figure}[htb]
\centering
\ig[width=.42\textwidth]{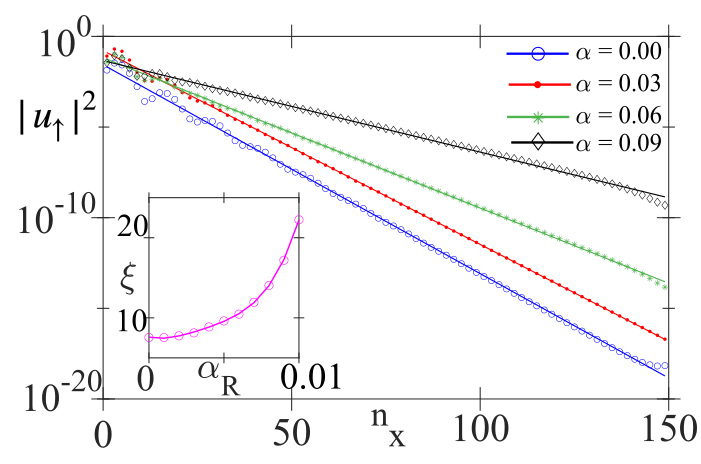}
\caption{Semi-log plot of the decay length of the "particle up" component $|u_\ua|^2$ 
of the edge-state spinor  as a function of position at momentum $k_y\approx0.75$ for different
values of Rashba SOC strength $\al_R$, while keeping all the other parameters unchanged.
Here the parameters are $m=1$, $\tilde{\mu} = -0.3$,
$\lam_I =0.15$, $h=0.1$ and $ \De = 0.06$.}
\label{fig:Decay}
\end{figure}

Figure \ref{fig:Decay} shows the spatial profile of the degenerate midgap states for a fixed $k_y =0.75$
between the pairs of nodes, for different values of Rashba spin-orbit (indicated by the markers). Here
$ n_x$ indicates the position along the chain. Solid lines indicate a best fit to an exponential law
$\psi(x) \approx \exp(-x/\xi)$, showing  that the states remain exponentially localized even for
finite $\alpha_R $. Importantly, unlike the $ \alpha_R= 0$ case, for finite $\alpha_R\neq 0 $ the edge
states do not satisfy the Majorana condition $\psi_b\neq(u_\uparrow,u_\downarrow,u_\uparrow^*,u_\downarrow^*)^T $.
This observation is also consistent with the analytic derivation of the boundary mode in the continuum limit,
see Appendix \ref{sec:analytics}.
The inset in Fig. \ref{fig:Decay} shows the best fit of the decay length with increasing Rashba SOC.

 \section{Josephson Junction} \label{sec:JJ}
\begin{figure}[htb]
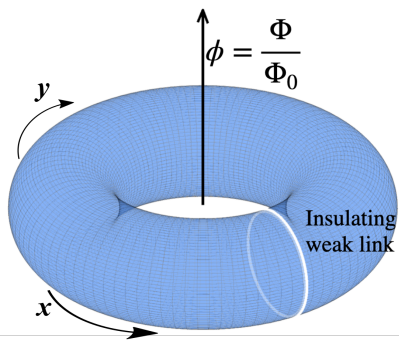

\centering
\ig[width=.3 \textwidth]{JJSchematic.png}
\caption{Schematic showing the torus-geometry used to study the Josephson junction. The y-direction has
periodic boundary conditions, making $k_y$ a good quantum number. The change in the phase of the
superconducting pairing corresponds to a flux $\Phi$ through this torus. The $1D$ chain along the $x-$direction
has twisted boundary conditions, i.e. all hopping parameters across the insulating weak link shown in white,
acquire a phase $\phi = \Phi/\Phi_0$, $\Phi_0$ being the flux quantum, and are attenuated by a factor
proportional to the strength of the insulating barrier}
\label{fig:JJtorus}
\end{figure}

To study the Josephson energy-phase relation we close the finite ribbon into a torus-like geometry
by adding a weak link between the first and last sites of the effective 1D chain, see Fig. \ref{fig:JJtorus}. 
All hopping terms across the weak link acquire a phase $\phi$ and are also attenuated by a factor proportional to
 the strength of the insulating barrier. 
Varying the phase difference $\phi$ across the junction
for a given $k_y$ allows to obtain the energy phase relation. 

\begin{figure}[htb]
\centering
\ig[width=.4\textwidth]{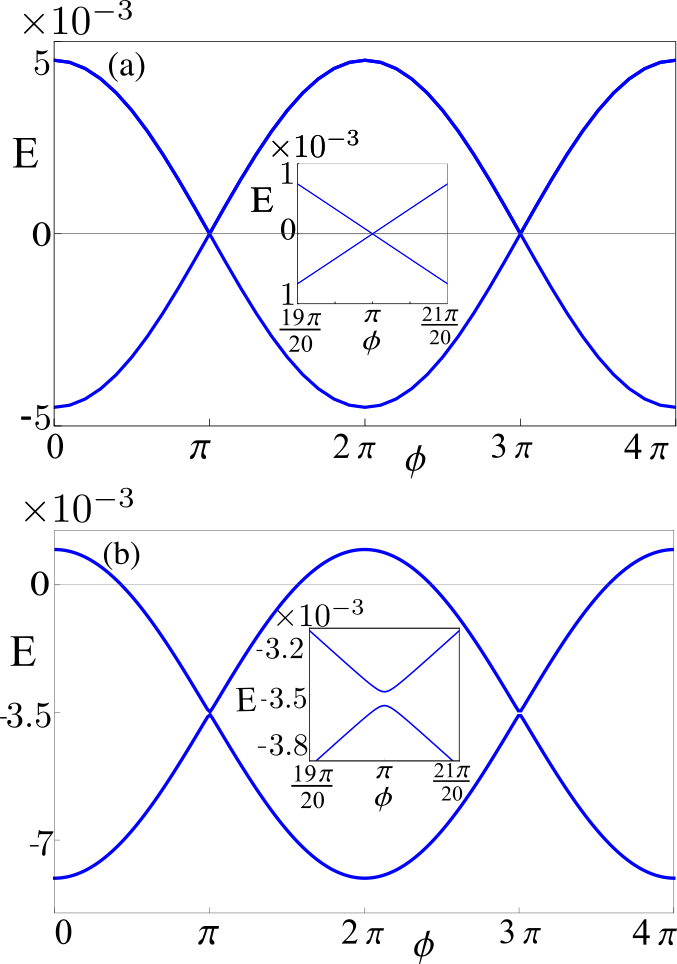}
\caption{Energy-phase relation of a Josephson junction  obtained
by solving the lattice model on a torus geometry of Fig. \ref{fig:JJtorus}. The Rashba SOC is
(a) $\al_R=0$ and (b) $\al_R=0.01$. We have chosen the momentum $k_y = 0.75$ which lies between
two nodal points. In (a) we find that there are zero-crossings at $\phi= \pi, 3\pi$ as is clear
from the inset. This means $E_J(\phi)$ has a $4\pi$ periodicity. However in (b) the Josephson
energy-phase relation is no longer symmetric about $E=0$. Additionally, there is no crossing at 
$\phi=\pi, 3\pi$ and therefore only a $2\pi$ periodicity in $E(\phi)$, the parameters are $m=1$, $\tilde{\mu} = -0.3$,
$\lam_I =0.15$, $h=0.1$ and $ \De = 0.06$}
\label{fig:JosephE}
\end{figure}

Fig. \ref{fig:JosephE} shows  $E(\phi)$ for the mid-gap states at a fixed $k_y =0.76 $ that lie between pair of
nodal points. For $\alpha_R =0$ this value of $k_y =0.76 $  corresponds to a topological non-trivial phase of
class BDI with winding $W=1 $. Conversely, with $ \alpha_R\neq 0 $ this $k_y $ value corresponds to a crystalline
topological phase with $\nu_{\cal M} = -1 $, see Fig. \ref{fig:inversion}.
In the absence of Rashba SOC, shown in Fig. \ref{fig:JosephE} (a),
the Josephson energy exhibits a $4\pi$ periodicity similar to the continuous model studied in
Ref. \onlinecite{sesh22}, with the energy levels crossing zero   
at $\phi=\pi$ and $3\pi$. This indicates the presence of Majorana edge states localized in the vicinity of the 
 weak link which decay exponentially into the bulk.  

When the Rashba SOC is finite,  shown in Fig. \ref{fig:JosephE} (b), we find that   $E(\phi)$ is no longer
symmetric about the $E=0$ line i.e. it is shifted away from zero. Moreover, an energy gap
opens at $\phi=\pi,~3\pi$ in the presence of a Rashba SOC as is clear from the inset in Fig.
\ref{fig:JosephE} (b).
Hence, in the presence of Rashba SOC the Josephson energy-phase
relation  has  a $2\pi$ periodicity for all  $k_y$ values. This is  consistent with the observation that the exponentially localized boundary states are not Majorana modes.


\section{Conclusion}
We have studied the effect of Rashba spin-orbit coupling on the nodal superconducting phase of an Ising
superconductor. This nodal phase was predicted in monolayer TMD's such as NbSe$_2$ in the presence of an
in-plane field which exceeds the Pauli limit $|h|>\Delta $ \cite{He2018,Fischer2018}.
The presence of Rashba SOC breaks the chiral symmetry and generally lifts the nodal points, resulting
in a fully gapped state. However, when the magnetic field is aligned along the $ \Gamma-K$ line the
system has a residual   mirror  symmetry $M_x $ which protects the nodal points at $ k_x=0$. The system therefore
realizes a nodal crystalline phase, characterized by   states exponentially localized at the (non-self-reflecting) x-boundary, which disperse parallel to the boundary, provided that the x-boundary preserves the
crystalline symmetry. 
However, we find that even in the presence of exponentially localized boundary states, the current
phase relation in a Josephson junction becomes trivial and follows a $ 2\pi $ periodicity. 

We note that
Rashba spin-orbit coupling is typically present in experimental setups and can be controlled using gates
and by changing substrates. This gives an experimental knob to tune in and out of the topological phase,
thus changing  the $4\pi  $ periodic current phase relation to the trivial $2\pi$. 

\section{Acknowledgements}
The authors would like to thank Hadar Steinberg and Ganapathy Murthy for fruitful discussions. 
D.M.  acknowledges support from the Israel Science Foundation (ISF) (grant No. 1884/18). 
M.K. and D.M. acknowledge support from the ISF, (grant No. 1251/19).

\appendix
 \section{Effective Hamiltonian for a $1D$ chain}\label{sec:ribbonH}
We study the tight binding lattice hamiltonian in a ribbon-like geometry with open boundary conditions in the $x $ direction and periodic boundary conditions in the  $y-$direction. Treating $k_y $ as a parameter, the resulting model describes a family of $1D$ chains along the $x$ direction. 
Below we setting the lattice parameter $a=1$.
The resulting family of 1D chains is described by  \eqref{eq:Hribbon} with the terms corresponding to kinetic energy $\mH_0$, Ising SOC $\mH_I$ and Rashba SOC $\mH_R$ are all dependent
on $k_y$, and involve terms which couple nearest as well as next nearest-neighbors:
\bea
H_{0}(k_y)&=&2t\cos(\sqrt{3}k_y)\sum_{i,\si} c^\dag_{i+1,\si}c_{i,\si} +{\rm h.c.} \non \\
&+&t \sum_{i,\si} c^\dag_{i+2,\sig}c_{i,\si} + {\rm h.c.}-\tilde{\mu} \sum_{i,\si} c^\dag_{i,\si}c_{i,\si} \\
H_{I}(k_y)&=&-2i\lam_I\cos(\sqrt{3}k_y)\sum_{i,\al,\be}c^\dag_{i+1,\al}\si^z_{\al\be}c_{i,\be}+{\rm h.c.} \non \\
&+& i\lam_I\sum_{i,\al,\be}  c^\dag_{i+2,\al}\si^z_{\al\be}c_{i,\be} + {\rm h.c.} \\
H_{R}(k_y)&=&-\frac{\al_R}{2\sqrt{3}}\sin(\sqrt{3}k_y)\sum_{i,\al,\be}c^\dag_{i+1,\al}\si^x_{\al\be}c_{i,\be}+{\rm h.c.}\non\\
&~& +i\frac {\al_R}{6} \cos(\sqrt{3}k_y) \sum_{i,\al,\be}\Big[c^\dag_{i+1,\al}\si^y_{\al\be}c_{i,\be} \non \\
&~& ~~~~~~~~~~~~~~~+ c^\dag_{i+2,\al}\si^y_{\al\be}c_{i,\be} + {\rm h.c.}\Big].
\eea

The in-plane magnetic field arises from on-site terms,
\bea
H_{B}=\sum_{i,\al,\be}\left(\bf{h}\cdot\bm{\si}\right)_{\al,\be}c^\dag_{i,\al}c_{i,\be}
\eea
Note that in all the terms above, we have suppressed the index $k_y$ for the creation (annihilation) operators.
However, since the superconducting term couples particle and hole components, it is written as 
\beq
H_{SC}=\sum_i\De c^\dag_{i,k_y\ua} c^\dag_{i,-k_y,\da}+{\rm h.c.}.
\eeq

The Hamiltonian in Eq. \eqref{eq:Hribbon} is used to obtain the excitation spectrum in Fig. \ref{fig:Decay} and \ref{fig:invbreak}, as well as the Josephson current phase relation Fig. \ref{fig:JosephE}.




\section{Derivation of boundary modes in the continuum limit of the  effective 1D model} \label{sec:analytics}
We consider the continuum limit of the effective 1D Hamiltonian obtained from \eqref{eq:Hkk} by treating $k_y $ as a parameter.  Following a similar analysis as in Ref. \onlinecite{klin21}, we focus on the regime of strong Ising spin-orbit coupling $\lambda_I\gg h,\Delta,\alpha_R,\mu_{k_y} $ where the magnetic field, superconductivity, and Rashba SOC can be treated as weak perturbations. We, therefore, consider initially the following bare Hamiltonian:
\beq\label{eq:HcFree}
    \mH_{0}^{1D}(k) =\frac{k^2}{2m}-\mu_{k_y}+\lam_I k(k^2-3k_y^2)\sig^z+\al_R (k\sig^y)
\eeq
setting aside the gap-opening terms such as magnetic field, SC, and the transverse Rashba term. Here $ k\equiv k_x$ is the momentum of the 1D system and the Pauli matrices $ \sigma$ operate on the spin basis.
For simplicity we consider the $k_y $ for which $ \mu_{k_y}=0$. 

\begin{figure}[htb]
    \centering  \includegraphics[width=.45\textwidth]{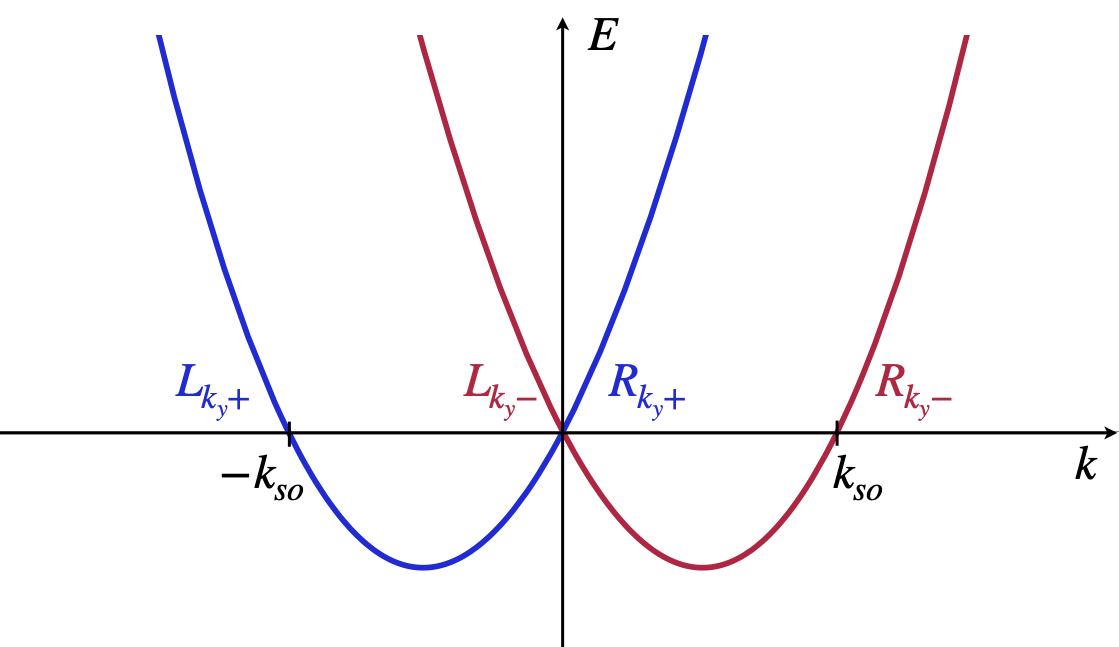}
    \caption{Spectrum of the bare 1D effective Hamiltonian Eq. \eqref{eq:HcFree} consists of two spin-orbit bands (marked by $\pm $) that cross the Fermi level $\mu_{k_y}=0 $ at the Fermi momenta $k_F =0,\pm k_{so}$.}
    \label{fig:bare1D}
\end{figure}

The eigenvalues of the bare 1D Hamiltonian \eqref{eq:HcFree} are given by:
\begin{eqnarray}
E(k)= \frac{k^2}{2m}\pm k\sqrt{\al_R ^2+\lam ^2 \left(k^2-3 k_y^2\right)^2}
\end{eqnarray}
and the Fermi points that satisfy  $ka\ll1 $ are located at $k=0 $ and $ k_{so}~=\pm\sqrt{3 k_y^2-\frac{\sqrt{16 \lam ^2 m^2 
\left(3 k_y^2-4 \al_R ^2 m^2\right)+1}-1}{8 \lam ^2 m^2}}$.
Fig. \ref{fig:bare1D} shows the spectrum of the bare Hamiltonian in the limit $ka\ll1 $. In addition, we have 3 gap-opening perturbations:
\begin{eqnarray}
    \mH_z &=& h \int_x\Psi_{k_y}(x)^\dag \sig^x\Psi_{k_y}\\
    \mH_\Delta&=& \Delta\int_x \Psi_{k_y}(x)i\sig^y  \Psi_{-k_y}(x)\\
    \mH_r &=& -\al_R k_y\int_x \Psi_{k_y}^\dag \sig^x \Psi_{k_y}
\end{eqnarray}
where we have introduced the spinor notation $\Psi_{k_y} =(\psi_{k_y,\uparrow},\psi_{k_y,\downarrow})^T $.
Next, we linearize the spectrum close to the Fermi points, see Fig. \ref{fig:bare1D}. The fields then take the form,
\begin{eqnarray}
    \Psi_{k_y+}(x) = e^{-i k_{so} x}L_{k_y+}(x)+R_{k_y+}(x)\\
    \Psi_{k_y-}(x) = e^{i k_{so} x}R_{k_y-}(x)+L_{k_y-}(x)
\end{eqnarray}
where $R_{k_y\si} (x)$ and $L_{k_y\si} (x)$ are slowly varying, and 
the spin-orbit eigenvectors of the bare Hamiltonian \eqref{eq:HcFree} are given by $ \Psi_{k_y-} = (i\sin \chi_k/2,\cos\chi_k/2)^T$,
$\Psi_{k_y+} = (\cos\chi_k/2,i\sin \chi_k/2)^T$,  with $b_k\cos\chi_k = \lam_I k(k^2-3k_y^2)$ and $b_k\sin\chi_k = \al_R k $.
Note that for strong Ising SOC the spins are aligned along the $z$ direction $\tan\chi_k \rightarrow 0 $. 

Ignoring strongly oscillatory terms, the kinetic energy can be written as 
\begin{eqnarray}
    \mH_{0}&=&-iv_i\! \int_x\! R_{k_y+}^\dag(x)\partial_x R_{k_y+}-L_{k_y-}^\dag(x)\partial_x L_{k_y-}\\
  &&  -iv_e \!\int_x\! R_{k_y-}^\dag(x)\partial_x R_{k_y-}-L_{k_y+}^\dag(x)\partial_x L_{k_y+}
\end{eqnarray}
and the gap-opening terms then become,
\begin{eqnarray}
    \mH_z &=& h \int_x R_{k_y+}^\dag(x) L_{k_y-}(x)+{\rm h.c.}\\
    \nonumber
    \mH_\Delta&=& \Delta\int_x \left[ R_{k_y+} (x) L_{-k_y-}(x)\right.\\&&+\left. L_{k_y+}(x) R_{-k_y-}(x)\right]\\
    \mH_r &=& -\al_Rk_y\int_x  R_{k_y+}^\dag (x)L_{k_y-}(x)+{\rm h.c.}.
\end{eqnarray}
The Hamiltonian separates into two decoupled subsystems which we label the ``external" (e) and ``internal"~(i)
branches $\Phi_e=~(L_{k_y+},R_{k_y-},L_{-k_y+}^\dag,R_{-k_y-}^\dag)^T $
and $\Phi_i=~(R_{k_y+},L_{k_y-},R_{-k_y+}^\dag,L_{-k_y-}^\dag)^T$ with the respective Hamiltonians:
\begin{eqnarray}
    \mH_i&=&-iv_i\partial_x \sig^z+h\tau^z\sig^x-\al_Rk_y\sig^x+\Delta\tau^y\sig^y\\
    \mH_e&=&iv_e\partial_x \sig^z+\Delta\tau^y\sig^y
\end{eqnarray}
In what follows we will drop the subscript $k_y $ for brevity.

We solve for a semi-infinite wire with a boundary at $x=0$. We make the following ansatz for the zero mode,
$\mH_{l}\phi_{l}(x) =0 $ with $\phi_{l}(x) = e^{-x/\xi_l} \phi_{l}(0)$
where $l=e/i $
with  $ \xi_e = v_e/ \Delta$ and two possible values $\xi_{i} $ for the inner branch  
$ \xi_{i 1} = \frac{v_i}{h+\sqrt{\Delta ^2+\al_R ^2 k_y^2}}$ and  $ \xi_{i 2} = \frac{v_i}{h-\sqrt{\Delta ^2+\al_R ^2 k_y^2}}$. 
Reincorporating the oscillatory phases and expressing the zero mode solutions in terms of the original basis  
${\bf \Psi} = (\Psi_{k_y+},\Psi_{k_y-},\Psi_{-k_y+}^\dag,\Psi_{-k_y-}^\dag)^T$ we find,
\begin{eqnarray}
\psi_{i1}= \phi_{i1} = e^{-x/\xi_{i1}} \left(\begin{array}{c}
        -i  \\
         -1\\
         \frac{i}{\be}\\
         -\frac{1}{\be}
    \end{array}\right) \\
  \psi_{i2}=  \phi_{i2} = e^{-x/\xi_{i2}} \left(\begin{array}{c}
        i  \\
         1\\
         +i\be\\
         -\be
    \end{array}\right) 
\end{eqnarray}
with $\be = \frac{\al_R  k_y-\sqrt{\Delta ^2+\al_R ^2 k_y^2}}{\Delta } $
and 
\begin{eqnarray}
    \psi_{e1} = e^{-x/\xi_{e}} 
    \left(\begin{array}{c}
        i e^{-ik_{so} x} \\
        e^{ik_{so} x}\\
         -ie^{ik_{so} x}\\
        e^{-ik_{so} x}
    \end{array}\right) \\
    \psi_{e2} = e^{-x/\xi_{e}}
    \left(\begin{array}{c}
       i  e^{-ik_{so} x} \\
         -e^{ik_{so} x}\\
         ie^{ik_{so} x}\\
         e^{-ik_{so} x}
    \end{array}\right) 
\end{eqnarray}
This allows us to construct a zero mode that satisfies the boundary conditions at $x=0$ namely $ \psi_M(x=0)=0$ which is:
\begin{eqnarray}
\nonumber
     \psi_M(x) &=&\frac{\be  (\be +1)}{\be -1}\psi_{i1}+\psi_{i2}+\frac{ (\be^2 +1)}{\be-1 }\psi_{e1}\\
  \nonumber
   &=& \frac{\be  (\be +1)}{\be -1} e^{-x/\xi_{i1}} \left(\begin{array}{c}
        -i  \\
         -1\\
         \frac{i}{\be}\\
         -\frac{1}{\be}
    \end{array}\right) + e^{-x/\xi_{i2}} \left(\begin{array}{c}
        i  \\
         1\\
         +i\be\\
         -\be
    \end{array}\right)\\& +&\frac{ (\be^2 +1)}{\be-1 }e^{-x/\xi_{e}} 
    \left(\begin{array}{c}
        i e^{-ik_{so} x} \\
        e^{ik_{so} x}\\
         -ie^{ik_{so} x}\\
        e^{-ik_{so} x}
    \end{array}\right) 
\end{eqnarray}
We note that in the absence of Rashba SOC, $\beta=-1 $, and the boundary mode indeed satisfies the Majorana condition namely $ \psi_M|_{\al_R=0}=(u_{+}(x),u_{-}(x),u_{+}^*(x),u_{-}^*(x))^T$. However, at finite $\al_R\neq0 $ this condition is no longer met and the boundary state is no longer a Majorana mode.

\textcolor{red}{}

 \section{Breaking of reflection symmetry}\label{sec:invbreak}
\begin{figure}[htb]
\begin{center}
\ig[width=.42\textwidth]{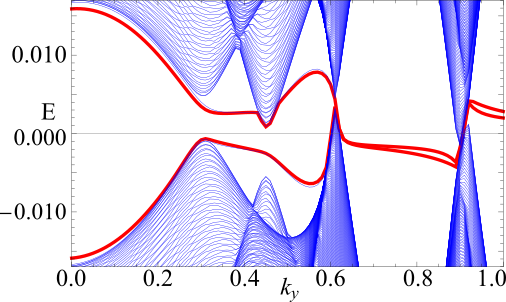}
\end{center}
\caption{Bulk and edge state spectrum for $\al=-0.01$, $\tilde{\mu}=-0.25$ $\mu_l=-0.175$, $m=1$,
$\lam_{SO} =0.15$, $h=0.1 $and $ \Delta = 0.06$. $\mu_l \neq 0$ on either of the two edges,
breaks the reflection symmetry and lifts the degeneracy of the edge states (shown as solid red lines).}
\label{fig:invbreak}
\end{figure}

In the family of 1D chains with open boundary conditions \eqref{eq:Hribbon}, the degeneracy of the mid-gap states
is protected by Mirror symmetry $M_x$,  i.e. the edge states  are reflected onto each
other under this symmetry. Breaking the  symmetry by adding a local  potential 
on one of the two edges $\mu_l \neq \tilde{\mu}$  will lift the degeneracy. This situation is shown in
Fig. \ref{fig:invbreak}.

Conversely, in the absence
of Rashba SOC, i.e. when $\al_R=0$ the  degeneracy is protected by the Chiral symmetry of the  1D chain for
fixed $k_y$. Consequently, the degeneracy is not lifted even in the presence of a local chemical potential.

\newpage

 


\bibliography{refs}

\begin{thebibliography}{43}%
\makeatletter
\providecommand \@ifxundefined [1]{%
 \@ifx{#1\undefined}
}%
\providecommand \@ifnum [1]{%
 \ifnum #1\expandafter \@firstoftwo
 \else \expandafter \@secondoftwo
 \fi
}%
\providecommand \@ifx [1]{%
 \ifx #1\expandafter \@firstoftwo
 \else \expandafter \@secondoftwo
 \fi
}%
\providecommand \natexlab [1]{#1}%
\providecommand \enquote  [1]{``#1''}%
\providecommand \bibnamefont  [1]{#1}%
\providecommand \bibfnamefont [1]{#1}%
\providecommand \citenamefont [1]{#1}%
\providecommand \href@noop [0]{\@secondoftwo}%
\providecommand \href [0]{\begingroup \@sanitize@url \@href}%
\providecommand \@href[1]{\@@startlink{#1}\@@href}%
\providecommand \@@href[1]{\endgroup#1\@@endlink}%
\providecommand \@sanitize@url [0]{\catcode `\\12\catcode `\$12\catcode
  `\&12\catcode `\#12\catcode `\^12\catcode `\_12\catcode `\%12\relax}%
\providecommand \@@startlink[1]{}%
\providecommand \@@endlink[0]{}%
\providecommand \url  [0]{\begingroup\@sanitize@url \@url }%
\providecommand \@url [1]{\endgroup\@href {#1}{\urlprefix }}%
\providecommand \urlprefix  [0]{URL }%
\providecommand \Eprint [0]{\href }%
\providecommand \doibase [0]{https://doi.org/}%
\providecommand \selectlanguage [0]{\@gobble}%
\providecommand \bibinfo  [0]{\@secondoftwo}%
\providecommand \bibfield  [0]{\@secondoftwo}%
\providecommand \translation [1]{[#1]}%
\providecommand \BibitemOpen [0]{}%
\providecommand \bibitemStop [0]{}%
\providecommand \bibitemNoStop [0]{.\EOS\space}%
\providecommand \EOS [0]{\spacefactor3000\relax}%
\providecommand \BibitemShut  [1]{\csname bibitem#1\endcsname}%
\let\auto@bib@innerbib\@empty
\bibitem [{\citenamefont {Zhou}\ \emph {et~al.}(2016)\citenamefont {Zhou},
  \citenamefont {Yuan}, \citenamefont {Jiang},\ and\ \citenamefont
  {Law}}]{Zhou2016}%
  \BibitemOpen
  \bibfield  {author} {\bibinfo {author} {\bibfnamefont {B.~T.}\ \bibnamefont
  {Zhou}}, \bibinfo {author} {\bibfnamefont {N.~F.~Q.}\ \bibnamefont {Yuan}},
  \bibinfo {author} {\bibfnamefont {H.-L.}\ \bibnamefont {Jiang}},\ and\
  \bibinfo {author} {\bibfnamefont {K.~T.}\ \bibnamefont {Law}},\ }\bibfield
  {title} {\bibinfo {title} {Ising superconductivity and majorana fermions in
  transition-metal dichalcogenides},\ }\href
  {https://doi.org/10.1103/PhysRevB.93.180501} {\bibfield  {journal} {\bibinfo
  {journal} {Physical Review B}\ }\textbf {\bibinfo {volume} {93}},\ \bibinfo
  {pages} {180501(R)} (\bibinfo {year} {2016})}\BibitemShut {NoStop}%
\bibitem [{\citenamefont {Ili{\'{c}}}\ \emph {et~al.}(2017)\citenamefont
  {Ili{\'{c}}}, \citenamefont {Meyer},\ and\ \citenamefont
  {Houzet}}]{Ilic2017}%
  \BibitemOpen
  \bibfield  {author} {\bibinfo {author} {\bibfnamefont {S.}~\bibnamefont
  {Ili{\'{c}}}}, \bibinfo {author} {\bibfnamefont {J.~S.}\ \bibnamefont
  {Meyer}},\ and\ \bibinfo {author} {\bibfnamefont {M.}~\bibnamefont
  {Houzet}},\ }\bibfield  {title} {\bibinfo {title} {{Enhancement of the Upper
  Critical Field in Disordered Transition Metal Dichalcogenide Monolayers}},\
  }\href {https://doi.org/10.1103/PhysRevLett.119.117001} {\bibfield  {journal}
  {\bibinfo  {journal} {Physical Review Letters}\ }\textbf {\bibinfo {volume}
  {119}},\ \bibinfo {pages} {117001} (\bibinfo {year} {2017})}\BibitemShut
  {NoStop}%
\bibitem [{\citenamefont {He}\ \emph {et~al.}(2018)\citenamefont {He},
  \citenamefont {Zhou}, \citenamefont {He}, \citenamefont {Yuan}, \citenamefont
  {Zhang},\ and\ \citenamefont {Law}}]{He2018}%
  \BibitemOpen
  \bibfield  {author} {\bibinfo {author} {\bibfnamefont {W.-Y.}\ \bibnamefont
  {He}}, \bibinfo {author} {\bibfnamefont {B.~T.}\ \bibnamefont {Zhou}},
  \bibinfo {author} {\bibfnamefont {J.~J.}\ \bibnamefont {He}}, \bibinfo
  {author} {\bibfnamefont {N.~i.~Q.}\ \bibnamefont {Yuan}}, \bibinfo {author}
  {\bibfnamefont {T.}~\bibnamefont {Zhang}},\ and\ \bibinfo {author}
  {\bibfnamefont {K.~T.}\ \bibnamefont {Law}},\ }\bibfield  {title} {\bibinfo
  {title} {Magnetic field driven nodal topological superconductivity in
  monolayer transition metal dichalcogenides},\ }\href
  {https://doi.org/10.1038/s42005-018-0041-4} {\bibfield  {journal} {\bibinfo
  {journal} {Communications Physics}\ }\textbf {\bibinfo {volume} {1}},\
  \bibinfo {pages} {40} (\bibinfo {year} {2018})}\BibitemShut {NoStop}%
\bibitem [{\citenamefont {Sosenko}\ \emph {et~al.}(2017)\citenamefont
  {Sosenko}, \citenamefont {Zhang},\ and\ \citenamefont {Aji}}]{Sosenko2017}%
  \BibitemOpen
  \bibfield  {author} {\bibinfo {author} {\bibfnamefont {E.}~\bibnamefont
  {Sosenko}}, \bibinfo {author} {\bibfnamefont {J.}~\bibnamefont {Zhang}},\
  and\ \bibinfo {author} {\bibfnamefont {V.}~\bibnamefont {Aji}},\ }\bibfield
  {title} {\bibinfo {title} {Unconventional superconductivity and anomalous
  response in hole-doped transition metal dichalcogenides},\ }\href
  {https://doi.org/10.1103/PhysRevB.95.144508} {\bibfield  {journal} {\bibinfo
  {journal} {Phys. Rev. B}\ }\textbf {\bibinfo {volume} {95}},\ \bibinfo
  {pages} {144508} (\bibinfo {year} {2017})}\BibitemShut {NoStop}%
\bibitem [{\citenamefont {Nakamura}\ and\ \citenamefont
  {Yanase}(2017)}]{Nakamura2017}%
  \BibitemOpen
  \bibfield  {author} {\bibinfo {author} {\bibfnamefont {Y.}~\bibnamefont
  {Nakamura}}\ and\ \bibinfo {author} {\bibfnamefont {Y.}~\bibnamefont
  {Yanase}},\ }\bibfield  {title} {\bibinfo {title} {Odd-parity
  superconductivity in bilayer transition metal dichalcogenides},\ }\href
  {https://doi.org/10.1103/PhysRevB.96.054501} {\bibfield  {journal} {\bibinfo
  {journal} {Phys. Rev. B}\ }\textbf {\bibinfo {volume} {96}},\ \bibinfo
  {pages} {054501} (\bibinfo {year} {2017})}\BibitemShut {NoStop}%
\bibitem [{\citenamefont {M{\"{o}}ckli}\ and\ \citenamefont
  {Khodas}(2018)}]{Mockli2018}%
  \BibitemOpen
  \bibfield  {author} {\bibinfo {author} {\bibfnamefont {D.}~\bibnamefont
  {M{\"{o}}ckli}}\ and\ \bibinfo {author} {\bibfnamefont {M.}~\bibnamefont
  {Khodas}},\ }\bibfield  {title} {\bibinfo {title} {{Robust parity-mixed
  superconductivity in disordered monolayer transition metal
  dichalcogenides}},\ }\href {https://doi.org/10.1103/PhysRevB.98.144518}
  {\bibfield  {journal} {\bibinfo  {journal} {Physical Review B}\ }\textbf
  {\bibinfo {volume} {98}},\ \bibinfo {pages} {144518} (\bibinfo {year}
  {2018})}\BibitemShut {NoStop}%
\bibitem [{\citenamefont {Fischer}\ \emph {et~al.}(2018)\citenamefont
  {Fischer}, \citenamefont {Sigrist},\ and\ \citenamefont
  {Agterberg}}]{Fischer2018}%
  \BibitemOpen
  \bibfield  {author} {\bibinfo {author} {\bibfnamefont {M.~H.}\ \bibnamefont
  {Fischer}}, \bibinfo {author} {\bibfnamefont {M.}~\bibnamefont {Sigrist}},\
  and\ \bibinfo {author} {\bibfnamefont {D.~F.}\ \bibnamefont {Agterberg}},\
  }\bibfield  {title} {\bibinfo {title} {Superconductivity without inversion
  and time-reversal symmetries},\ }\href
  {https://doi.org/10.1103/PhysRevLett.121.157003} {\bibfield  {journal}
  {\bibinfo  {journal} {Phys. Rev. Lett.}\ }\textbf {\bibinfo {volume} {121}},\
  \bibinfo {pages} {157003} (\bibinfo {year} {2018})}\BibitemShut {NoStop}%
\bibitem [{\citenamefont {Smidman}\ \emph
  {et~al.}(2017{\natexlab{a}})\citenamefont {Smidman}, \citenamefont {Salamon},
  \citenamefont {Yuan},\ and\ \citenamefont {Agterberg}}]{Smidman2017}%
  \BibitemOpen
  \bibfield  {author} {\bibinfo {author} {\bibfnamefont {M.}~\bibnamefont
  {Smidman}}, \bibinfo {author} {\bibfnamefont {M.~B.}\ \bibnamefont
  {Salamon}}, \bibinfo {author} {\bibfnamefont {H.~Q.}\ \bibnamefont {Yuan}},\
  and\ \bibinfo {author} {\bibfnamefont {D.~F.}\ \bibnamefont {Agterberg}},\
  }\bibfield  {title} {\bibinfo {title} {Superconductivity and spin-orbit
  coupling in non-centrosymmetric materials: a review},\ }\href
  {https://doi.org/10.1088/1361-6633/80/3/036501} {\bibfield  {journal}
  {\bibinfo  {journal} {Reports on Progress in Physics}\ }\textbf {\bibinfo
  {volume} {80}},\ \bibinfo {pages} {036501} (\bibinfo {year}
  {2017}{\natexlab{a}})}\BibitemShut {NoStop}%
\bibitem [{\citenamefont {Yuan}\ \emph {et~al.}(2014)\citenamefont {Yuan},
  \citenamefont {Mak},\ and\ \citenamefont {Law}}]{Yuan2014}%
  \BibitemOpen
  \bibfield  {author} {\bibinfo {author} {\bibfnamefont {N.~F.~Q.}\
  \bibnamefont {Yuan}}, \bibinfo {author} {\bibfnamefont {K.~F.}\ \bibnamefont
  {Mak}},\ and\ \bibinfo {author} {\bibfnamefont {K.~T.}\ \bibnamefont {Law}},\
  }\bibfield  {title} {\bibinfo {title} {Possible topological superconducting
  phases of ${\mathrm{mos}}_{2}$},\ }\href
  {https://doi.org/10.1103/PhysRevLett.113.097001} {\bibfield  {journal}
  {\bibinfo  {journal} {Phys. Rev. Lett.}\ }\textbf {\bibinfo {volume} {113}},\
  \bibinfo {pages} {097001} (\bibinfo {year} {2014})}\BibitemShut {NoStop}%
\bibitem [{\citenamefont {Oiwa}\ \emph {et~al.}(2018)\citenamefont {Oiwa},
  \citenamefont {Yanagi},\ and\ \citenamefont {Kusunose}}]{Oiwa2018}%
  \BibitemOpen
  \bibfield  {author} {\bibinfo {author} {\bibfnamefont {R.}~\bibnamefont
  {Oiwa}}, \bibinfo {author} {\bibfnamefont {Y.}~\bibnamefont {Yanagi}},\ and\
  \bibinfo {author} {\bibfnamefont {H.}~\bibnamefont {Kusunose}},\ }\bibfield
  {title} {\bibinfo {title} {Theory of superconductivity in hole-doped
  monolayer ${\mathrm{mos}}_{2}$},\ }\href
  {https://doi.org/10.1103/PhysRevB.98.064509} {\bibfield  {journal} {\bibinfo
  {journal} {Phys. Rev. B}\ }\textbf {\bibinfo {volume} {98}},\ \bibinfo
  {pages} {064509} (\bibinfo {year} {2018})}\BibitemShut {NoStop}%
\bibitem [{\citenamefont {Hsu}\ \emph {et~al.}(2017)\citenamefont {Hsu},
  \citenamefont {Vaezi}, \citenamefont {Fischer},\ and\ \citenamefont
  {Kim}}]{Hsu2017}%
  \BibitemOpen
  \bibfield  {author} {\bibinfo {author} {\bibfnamefont {Y.-T.}\ \bibnamefont
  {Hsu}}, \bibinfo {author} {\bibfnamefont {A.}~\bibnamefont {Vaezi}}, \bibinfo
  {author} {\bibfnamefont {M.~H.}\ \bibnamefont {Fischer}},\ and\ \bibinfo
  {author} {\bibfnamefont {E.-A.}\ \bibnamefont {Kim}},\ }\bibfield  {title}
  {\bibinfo {title} {Topological superconductivity in monolayer transition
  metal dichalcogenides},\ }\href {https://doi.org/10.1038/ncomms14985}
  {\bibfield  {journal} {\bibinfo  {journal} {Nature Communications}\ }\textbf
  {\bibinfo {volume} {8}},\ \bibinfo {pages} {14985} (\bibinfo {year}
  {2017})}\BibitemShut {NoStop}%
\bibitem [{\citenamefont {Wang}\ \emph {et~al.}(2018)\citenamefont {Wang},
  \citenamefont {Rosdahl},\ and\ \citenamefont {Sticlet}}]{Wang2018}%
  \BibitemOpen
  \bibfield  {author} {\bibinfo {author} {\bibfnamefont {L.}~\bibnamefont
  {Wang}}, \bibinfo {author} {\bibfnamefont {T.~O.}\ \bibnamefont {Rosdahl}},\
  and\ \bibinfo {author} {\bibfnamefont {D.}~\bibnamefont {Sticlet}},\
  }\bibfield  {title} {\bibinfo {title} {Platform for nodal topological
  superconductors in monolayer molybdenum dichalcogenides},\ }\href
  {https://doi.org/10.1103/PhysRevB.98.205411} {\bibfield  {journal} {\bibinfo
  {journal} {Phys. Rev. B}\ }\textbf {\bibinfo {volume} {98}},\ \bibinfo
  {pages} {205411} (\bibinfo {year} {2018})}\BibitemShut {NoStop}%
\bibitem [{\citenamefont {Oiwa}\ \emph {et~al.}(2019)\citenamefont {Oiwa},
  \citenamefont {Yanagi},\ and\ \citenamefont {Kusunose}}]{Oiwa2019}%
  \BibitemOpen
  \bibfield  {author} {\bibinfo {author} {\bibfnamefont {R.}~\bibnamefont
  {Oiwa}}, \bibinfo {author} {\bibfnamefont {Y.}~\bibnamefont {Yanagi}},\ and\
  \bibinfo {author} {\bibfnamefont {H.}~\bibnamefont {Kusunose}},\ }\bibfield
  {title} {\bibinfo {title} {Time-reversal symmetry breaking superconductivity
  in hole-doped monolayer mos2},\ }\href
  {https://doi.org/10.7566/JPSJ.88.063703} {\bibfield  {journal} {\bibinfo
  {journal} {Journal of the Physical Society of Japan}\ }\textbf {\bibinfo
  {volume} {88}},\ \bibinfo {pages} {063703} (\bibinfo {year} {2019})},\
  \Eprint {https://arxiv.org/abs/https://doi.org/10.7566/JPSJ.88.063703}
  {https://doi.org/10.7566/JPSJ.88.063703} \BibitemShut {NoStop}%
\bibitem [{\citenamefont {Sohn}\ \emph {et~al.}(2018)\citenamefont {Sohn},
  \citenamefont {Xi}, \citenamefont {He}, \citenamefont {Jiang}, \citenamefont
  {Wang}, \citenamefont {Kang}, \citenamefont {Park}, \citenamefont {Berger},
  \citenamefont {Forr{\'{o}}}, \citenamefont {Law}, \citenamefont {Shan},\ and\
  \citenamefont {Mak}}]{Sohn2018}%
  \BibitemOpen
  \bibfield  {author} {\bibinfo {author} {\bibfnamefont {E.}~\bibnamefont
  {Sohn}}, \bibinfo {author} {\bibfnamefont {X.}~\bibnamefont {Xi}}, \bibinfo
  {author} {\bibfnamefont {W.-Y.}\ \bibnamefont {He}}, \bibinfo {author}
  {\bibfnamefont {S.}~\bibnamefont {Jiang}}, \bibinfo {author} {\bibfnamefont
  {Z.}~\bibnamefont {Wang}}, \bibinfo {author} {\bibfnamefont {K.}~\bibnamefont
  {Kang}}, \bibinfo {author} {\bibfnamefont {J.-H.}\ \bibnamefont {Park}},
  \bibinfo {author} {\bibfnamefont {H.}~\bibnamefont {Berger}}, \bibinfo
  {author} {\bibfnamefont {L.}~\bibnamefont {Forr{\'{o}}}}, \bibinfo {author}
  {\bibfnamefont {K.~T.}\ \bibnamefont {Law}}, \bibinfo {author} {\bibfnamefont
  {J.}~\bibnamefont {Shan}},\ and\ \bibinfo {author} {\bibfnamefont {K.~F.}\
  \bibnamefont {Mak}},\ }\bibfield  {title} {\bibinfo {title} {{An unusual
  continuous paramagnetic-limited superconducting phase transition in 2D NbSe
  2}},\ }\href {https://doi.org/10.1038/s41563-018-0061-1} {\bibfield
  {journal} {\bibinfo  {journal} {Nature Materials}\ }\textbf {\bibinfo
  {volume} {17}},\ \bibinfo {pages} {504} (\bibinfo {year} {2018})}\BibitemShut
  {NoStop}%
\bibitem [{\citenamefont {Wang}\ \emph {et~al.}(2012)\citenamefont {Wang},
  \citenamefont {Kalantar-Zadeh}, \citenamefont {Kis}, \citenamefont
  {Coleman},\ and\ \citenamefont {Strano}}]{Wang2012}%
  \BibitemOpen
  \bibfield  {author} {\bibinfo {author} {\bibfnamefont {Q.~H.}\ \bibnamefont
  {Wang}}, \bibinfo {author} {\bibfnamefont {K.}~\bibnamefont
  {Kalantar-Zadeh}}, \bibinfo {author} {\bibfnamefont {A.}~\bibnamefont {Kis}},
  \bibinfo {author} {\bibfnamefont {J.~N.}\ \bibnamefont {Coleman}},\ and\
  \bibinfo {author} {\bibfnamefont {M.~S.}\ \bibnamefont {Strano}},\ }\bibfield
   {title} {\bibinfo {title} {Electronics and optoelectronics of
  two-dimensional transition metal dichalcogenides},\ }\href
  {https://doi.org/10.1038/nnano.2012.193} {\bibfield  {journal} {\bibinfo
  {journal} {Nature Nanotechnology}\ }\textbf {\bibinfo {volume} {7}},\
  \bibinfo {pages} {699} (\bibinfo {year} {2012})}\BibitemShut {NoStop}%
\bibitem [{\citenamefont {Geim}\ and\ \citenamefont
  {Grigorieva}(2013)}]{Geim2013}%
  \BibitemOpen
  \bibfield  {author} {\bibinfo {author} {\bibfnamefont {A.~K.}\ \bibnamefont
  {Geim}}\ and\ \bibinfo {author} {\bibfnamefont {I.~V.}\ \bibnamefont
  {Grigorieva}},\ }\bibfield  {title} {\bibinfo {title} {{Van der Waals
  heterostructures}},\ }\href {https://doi.org/10.1038/nature12385} {\bibfield
  {journal} {\bibinfo  {journal} {Nature}\ }\textbf {\bibinfo {volume} {499}},\
  \bibinfo {pages} {419} (\bibinfo {year} {2013})}\BibitemShut {NoStop}%
\bibitem [{\citenamefont {Lu}\ \emph {et~al.}(2015)\citenamefont {Lu},
  \citenamefont {Zheliuk}, \citenamefont {Leermakers}, \citenamefont {Yuan},
  \citenamefont {Zeitler}, \citenamefont {Law},\ and\ \citenamefont
  {Ye}}]{Lu2015}%
  \BibitemOpen
  \bibfield  {author} {\bibinfo {author} {\bibfnamefont {J.~M.}\ \bibnamefont
  {Lu}}, \bibinfo {author} {\bibfnamefont {O.}~\bibnamefont {Zheliuk}},
  \bibinfo {author} {\bibfnamefont {I.}~\bibnamefont {Leermakers}}, \bibinfo
  {author} {\bibfnamefont {N.~F.~Q.}\ \bibnamefont {Yuan}}, \bibinfo {author}
  {\bibfnamefont {U.}~\bibnamefont {Zeitler}}, \bibinfo {author} {\bibfnamefont
  {K.~T.}\ \bibnamefont {Law}},\ and\ \bibinfo {author} {\bibfnamefont {J.~T.}\
  \bibnamefont {Ye}},\ }\bibfield  {title} {\bibinfo {title} {{Evidence for
  two-dimensional Ising superconductivity in gated MoS2}},\ }\href
  {https://doi.org/10.1126/science.aab2277} {\bibfield  {journal} {\bibinfo
  {journal} {Science}\ }\textbf {\bibinfo {volume} {350}},\ \bibinfo {pages}
  {1353} (\bibinfo {year} {2015})}\BibitemShut {NoStop}%
\bibitem [{\citenamefont {Ugeda}\ \emph {et~al.}(2016)\citenamefont {Ugeda},
  \citenamefont {Bradley}, \citenamefont {Zhang}, \citenamefont {Onishi},
  \citenamefont {Chen}, \citenamefont {Ruan}, \citenamefont
  {Ojeda-Aristizabal}, \citenamefont {Ryu}, \citenamefont {Edmonds},
  \citenamefont {Tsai}, \citenamefont {Riss}, \citenamefont {Mo}, \citenamefont
  {Lee}, \citenamefont {Zettl}, \citenamefont {Hussain}, \citenamefont {Shen},\
  and\ \citenamefont {Crommie}}]{Ugeda2015}%
  \BibitemOpen
  \bibfield  {author} {\bibinfo {author} {\bibfnamefont {M.~M.}\ \bibnamefont
  {Ugeda}}, \bibinfo {author} {\bibfnamefont {A.~J.}\ \bibnamefont {Bradley}},
  \bibinfo {author} {\bibfnamefont {Y.}~\bibnamefont {Zhang}}, \bibinfo
  {author} {\bibfnamefont {S.}~\bibnamefont {Onishi}}, \bibinfo {author}
  {\bibfnamefont {Y.}~\bibnamefont {Chen}}, \bibinfo {author} {\bibfnamefont
  {W.}~\bibnamefont {Ruan}}, \bibinfo {author} {\bibfnamefont {C.}~\bibnamefont
  {Ojeda-Aristizabal}}, \bibinfo {author} {\bibfnamefont {H.}~\bibnamefont
  {Ryu}}, \bibinfo {author} {\bibfnamefont {M.~T.}\ \bibnamefont {Edmonds}},
  \bibinfo {author} {\bibfnamefont {H.-Z.}\ \bibnamefont {Tsai}}, \bibinfo
  {author} {\bibfnamefont {A.}~\bibnamefont {Riss}}, \bibinfo {author}
  {\bibfnamefont {S.-K.}\ \bibnamefont {Mo}}, \bibinfo {author} {\bibfnamefont
  {D.}~\bibnamefont {Lee}}, \bibinfo {author} {\bibfnamefont {A.}~\bibnamefont
  {Zettl}}, \bibinfo {author} {\bibfnamefont {Z.}~\bibnamefont {Hussain}},
  \bibinfo {author} {\bibfnamefont {Z.-X.}\ \bibnamefont {Shen}},\ and\
  \bibinfo {author} {\bibfnamefont {M.~F.}\ \bibnamefont {Crommie}},\
  }\bibfield  {title} {\bibinfo {title} {{Characterization of collective ground
  states in single-layer NbSe2}},\ }\href {https://doi.org/10.1038/nphys3527}
  {\bibfield  {journal} {\bibinfo  {journal} {Nature Physics}\ }\textbf
  {\bibinfo {volume} {12}},\ \bibinfo {pages} {92} (\bibinfo {year}
  {2016})}\BibitemShut {NoStop}%
\bibitem [{\citenamefont {Saito}\ \emph {et~al.}(2016)\citenamefont {Saito},
  \citenamefont {Nakamura}, \citenamefont {Bahramy}, \citenamefont {Kohama},
  \citenamefont {Ye}, \citenamefont {Kasahara}, \citenamefont {Nakagawa},
  \citenamefont {Onga}, \citenamefont {Tokunaga}, \citenamefont {Nojima},
  \citenamefont {Yanase},\ and\ \citenamefont {Iwasa}}]{Saito2016}%
  \BibitemOpen
  \bibfield  {author} {\bibinfo {author} {\bibfnamefont {Y.}~\bibnamefont
  {Saito}}, \bibinfo {author} {\bibfnamefont {Y.}~\bibnamefont {Nakamura}},
  \bibinfo {author} {\bibfnamefont {M.~S.}\ \bibnamefont {Bahramy}}, \bibinfo
  {author} {\bibfnamefont {Y.}~\bibnamefont {Kohama}}, \bibinfo {author}
  {\bibfnamefont {J.}~\bibnamefont {Ye}}, \bibinfo {author} {\bibfnamefont
  {Y.}~\bibnamefont {Kasahara}}, \bibinfo {author} {\bibfnamefont
  {Y.}~\bibnamefont {Nakagawa}}, \bibinfo {author} {\bibfnamefont
  {M.}~\bibnamefont {Onga}}, \bibinfo {author} {\bibfnamefont {M.}~\bibnamefont
  {Tokunaga}}, \bibinfo {author} {\bibfnamefont {T.}~\bibnamefont {Nojima}},
  \bibinfo {author} {\bibfnamefont {Y.}~\bibnamefont {Yanase}},\ and\ \bibinfo
  {author} {\bibfnamefont {Y.}~\bibnamefont {Iwasa}},\ }\bibfield  {title}
  {\bibinfo {title} {Superconductivity protected by spin-valley locking in
  ion-gated mos2},\ }\href {https://doi.org/10.1038/nphys3580} {\bibfield
  {journal} {\bibinfo  {journal} {Nature Physics}\ }\textbf {\bibinfo {volume}
  {12}},\ \bibinfo {pages} {144} (\bibinfo {year} {2016})}\BibitemShut
  {NoStop}%
\bibitem [{\citenamefont {Xi}\ \emph {et~al.}(2016)\citenamefont {Xi},
  \citenamefont {Wang}, \citenamefont {Zhao}, \citenamefont {Park},
  \citenamefont {Law}, \citenamefont {Berger}, \citenamefont {Forr{\'o}},
  \citenamefont {Shan},\ and\ \citenamefont {Mak}}]{Xi2016}%
  \BibitemOpen
  \bibfield  {author} {\bibinfo {author} {\bibfnamefont {X.}~\bibnamefont
  {Xi}}, \bibinfo {author} {\bibfnamefont {Z.}~\bibnamefont {Wang}}, \bibinfo
  {author} {\bibfnamefont {W.}~\bibnamefont {Zhao}}, \bibinfo {author}
  {\bibfnamefont {J.-H.}\ \bibnamefont {Park}}, \bibinfo {author}
  {\bibfnamefont {K.~T.}\ \bibnamefont {Law}}, \bibinfo {author} {\bibfnamefont
  {H.}~\bibnamefont {Berger}}, \bibinfo {author} {\bibfnamefont
  {L.}~\bibnamefont {Forr{\'o}}}, \bibinfo {author} {\bibfnamefont
  {J.}~\bibnamefont {Shan}},\ and\ \bibinfo {author} {\bibfnamefont {K.~F.}\
  \bibnamefont {Mak}},\ }\bibfield  {title} {\bibinfo {title} {Ising pairing in
  superconducting nbse2 atomic layers},\ }\href
  {https://doi.org/10.1038/nphys3538} {\bibfield  {journal} {\bibinfo
  {journal} {Nature Physics}\ }\textbf {\bibinfo {volume} {12}},\ \bibinfo
  {pages} {139} (\bibinfo {year} {2016})}\BibitemShut {NoStop}%
\bibitem [{\citenamefont {Costanzo}\ \emph {et~al.}(2016)\citenamefont
  {Costanzo}, \citenamefont {Jo}, \citenamefont {Berger},\ and\ \citenamefont
  {Morpurgo}}]{Costanzo2016}%
  \BibitemOpen
  \bibfield  {author} {\bibinfo {author} {\bibfnamefont {D.}~\bibnamefont
  {Costanzo}}, \bibinfo {author} {\bibfnamefont {S.}~\bibnamefont {Jo}},
  \bibinfo {author} {\bibfnamefont {H.}~\bibnamefont {Berger}},\ and\ \bibinfo
  {author} {\bibfnamefont {A.~F.}\ \bibnamefont {Morpurgo}},\ }\bibfield
  {title} {\bibinfo {title} {Gate-induced superconductivity in atomically thin
  mos2 crystals},\ }\href {https://doi.org/10.1038/nnano.2015.314} {\bibfield
  {journal} {\bibinfo  {journal} {Nature Nanotechnology}\ }\textbf {\bibinfo
  {volume} {11}},\ \bibinfo {pages} {339} (\bibinfo {year} {2016})}\BibitemShut
  {NoStop}%
\bibitem [{\citenamefont {Dvir}\ \emph {et~al.}(2018)\citenamefont {Dvir},
  \citenamefont {Massee}, \citenamefont {Attias}, \citenamefont {Khodas},
  \citenamefont {Aprili}, \citenamefont {Quay},\ and\ \citenamefont
  {Steinberg}}]{Dvir2017}%
  \BibitemOpen
  \bibfield  {author} {\bibinfo {author} {\bibfnamefont {T.}~\bibnamefont
  {Dvir}}, \bibinfo {author} {\bibfnamefont {F.}~\bibnamefont {Massee}},
  \bibinfo {author} {\bibfnamefont {L.}~\bibnamefont {Attias}}, \bibinfo
  {author} {\bibfnamefont {M.}~\bibnamefont {Khodas}}, \bibinfo {author}
  {\bibfnamefont {M.}~\bibnamefont {Aprili}}, \bibinfo {author} {\bibfnamefont
  {C.~H.~L.}\ \bibnamefont {Quay}},\ and\ \bibinfo {author} {\bibfnamefont
  {H.}~\bibnamefont {Steinberg}},\ }\bibfield  {title} {\bibinfo {title}
  {{Spectroscopy of bulk and few-layer superconducting NbSe2 with van der Waals
  tunnel junctions}},\ }\href {https://doi.org/10.1038/s41467-018-03000-w}
  {\bibfield  {journal} {\bibinfo  {journal} {Nature Communications}\ }\textbf
  {\bibinfo {volume} {9}},\ \bibinfo {pages} {598} (\bibinfo {year}
  {2018})}\BibitemShut {NoStop}%
\bibitem [{\citenamefont {de~la Barrera}\ \emph {et~al.}(2018)\citenamefont
  {de~la Barrera}, \citenamefont {Sinko}, \citenamefont {Gopalan},
  \citenamefont {Sivadas}, \citenamefont {Seyler}, \citenamefont {Watanabe},
  \citenamefont {Taniguchi}, \citenamefont {Tsen}, \citenamefont {Xu},
  \citenamefont {Xiao},\ and\ \citenamefont {Hunt}}]{DelaBarrera2018}%
  \BibitemOpen
  \bibfield  {author} {\bibinfo {author} {\bibfnamefont {S.~C.}\ \bibnamefont
  {de~la Barrera}}, \bibinfo {author} {\bibfnamefont {M.~R.}\ \bibnamefont
  {Sinko}}, \bibinfo {author} {\bibfnamefont {D.~P.}\ \bibnamefont {Gopalan}},
  \bibinfo {author} {\bibfnamefont {N.}~\bibnamefont {Sivadas}}, \bibinfo
  {author} {\bibfnamefont {K.~L.}\ \bibnamefont {Seyler}}, \bibinfo {author}
  {\bibfnamefont {K.}~\bibnamefont {Watanabe}}, \bibinfo {author}
  {\bibfnamefont {T.}~\bibnamefont {Taniguchi}}, \bibinfo {author}
  {\bibfnamefont {A.~W.}\ \bibnamefont {Tsen}}, \bibinfo {author}
  {\bibfnamefont {X.}~\bibnamefont {Xu}}, \bibinfo {author} {\bibfnamefont
  {D.}~\bibnamefont {Xiao}},\ and\ \bibinfo {author} {\bibfnamefont {B.~M.}\
  \bibnamefont {Hunt}},\ }\bibfield  {title} {\bibinfo {title} {Tuning ising
  superconductivity with layer and spin-orbit coupling in two-dimensional
  transition-metal dichalcogenides},\ }\href
  {https://doi.org/10.1038/s41467-018-03888-4} {\bibfield  {journal} {\bibinfo
  {journal} {Nature Communications}\ }\textbf {\bibinfo {volume} {9}},\
  \bibinfo {pages} {1427} (\bibinfo {year} {2018})}\BibitemShut {NoStop}%
\bibitem [{\citenamefont {Xing}\ \emph {et~al.}(2017)\citenamefont {Xing},
  \citenamefont {Zhao}, \citenamefont {Shan}, \citenamefont {Zheng},
  \citenamefont {Zhang}, \citenamefont {Fu}, \citenamefont {Liu}, \citenamefont
  {Tian}, \citenamefont {Xi}, \citenamefont {Liu}, \citenamefont {Feng},
  \citenamefont {Lin}, \citenamefont {Ji}, \citenamefont {Chen}, \citenamefont
  {Xue},\ and\ \citenamefont {Wang}}]{Xing2017}%
  \BibitemOpen
  \bibfield  {author} {\bibinfo {author} {\bibfnamefont {Y.}~\bibnamefont
  {Xing}}, \bibinfo {author} {\bibfnamefont {K.}~\bibnamefont {Zhao}}, \bibinfo
  {author} {\bibfnamefont {P.}~\bibnamefont {Shan}}, \bibinfo {author}
  {\bibfnamefont {F.}~\bibnamefont {Zheng}}, \bibinfo {author} {\bibfnamefont
  {Y.}~\bibnamefont {Zhang}}, \bibinfo {author} {\bibfnamefont
  {H.}~\bibnamefont {Fu}}, \bibinfo {author} {\bibfnamefont {Y.}~\bibnamefont
  {Liu}}, \bibinfo {author} {\bibfnamefont {M.}~\bibnamefont {Tian}}, \bibinfo
  {author} {\bibfnamefont {C.}~\bibnamefont {Xi}}, \bibinfo {author}
  {\bibfnamefont {H.}~\bibnamefont {Liu}}, \bibinfo {author} {\bibfnamefont
  {J.}~\bibnamefont {Feng}}, \bibinfo {author} {\bibfnamefont {X.}~\bibnamefont
  {Lin}}, \bibinfo {author} {\bibfnamefont {S.}~\bibnamefont {Ji}}, \bibinfo
  {author} {\bibfnamefont {X.}~\bibnamefont {Chen}}, \bibinfo {author}
  {\bibfnamefont {Q.-K.}\ \bibnamefont {Xue}},\ and\ \bibinfo {author}
  {\bibfnamefont {J.}~\bibnamefont {Wang}},\ }\bibfield  {title} {\bibinfo
  {title} {Ising superconductivity and quantum phase transition in macro-size
  monolayer nbse2},\ }\href {https://doi.org/10.1021/acs.nanolett.7b03026}
  {\bibfield  {journal} {\bibinfo  {journal} {Nano Letters}\ }\textbf {\bibinfo
  {volume} {17}},\ \bibinfo {pages} {6802} (\bibinfo {year}
  {2017})}\BibitemShut {NoStop}%
\bibitem [{\citenamefont {Hamill}\ \emph {et~al.}(2021)\citenamefont {Hamill},
  \citenamefont {Heischmidt}, \citenamefont {Sohn}, \citenamefont {Shaffer},
  \citenamefont {Tsai}, \citenamefont {Zhang}, \citenamefont {Xi},
  \citenamefont {Suslov}, \citenamefont {Berger}, \citenamefont {Forr{\'o}},
  \citenamefont {Burnell}, \citenamefont {Shan}, \citenamefont {Mak},
  \citenamefont {Fernandes}, \citenamefont {Wang},\ and\ \citenamefont
  {Pribiag}}]{Hamill2021}%
  \BibitemOpen
  \bibfield  {author} {\bibinfo {author} {\bibfnamefont {A.}~\bibnamefont
  {Hamill}}, \bibinfo {author} {\bibfnamefont {B.}~\bibnamefont {Heischmidt}},
  \bibinfo {author} {\bibfnamefont {E.}~\bibnamefont {Sohn}}, \bibinfo {author}
  {\bibfnamefont {D.}~\bibnamefont {Shaffer}}, \bibinfo {author} {\bibfnamefont
  {K.-T.}\ \bibnamefont {Tsai}}, \bibinfo {author} {\bibfnamefont
  {X.}~\bibnamefont {Zhang}}, \bibinfo {author} {\bibfnamefont
  {X.}~\bibnamefont {Xi}}, \bibinfo {author} {\bibfnamefont {A.}~\bibnamefont
  {Suslov}}, \bibinfo {author} {\bibfnamefont {H.}~\bibnamefont {Berger}},
  \bibinfo {author} {\bibfnamefont {L.}~\bibnamefont {Forr{\'o}}}, \bibinfo
  {author} {\bibfnamefont {F.~J.}\ \bibnamefont {Burnell}}, \bibinfo {author}
  {\bibfnamefont {J.}~\bibnamefont {Shan}}, \bibinfo {author} {\bibfnamefont
  {K.~F.}\ \bibnamefont {Mak}}, \bibinfo {author} {\bibfnamefont {R.~M.}\
  \bibnamefont {Fernandes}}, \bibinfo {author} {\bibfnamefont {K.}~\bibnamefont
  {Wang}},\ and\ \bibinfo {author} {\bibfnamefont {V.~S.}\ \bibnamefont
  {Pribiag}},\ }\bibfield  {title} {\bibinfo {title} {Two-fold symmetric
  superconductivity in few-layer nbse2},\ }\bibfield  {journal} {\bibinfo
  {journal} {Nature Physics}\ }\href
  {https://doi.org/10.1038/s41567-021-01219-x} {10.1038/s41567-021-01219-x}
  (\bibinfo {year} {2021})\BibitemShut {NoStop}%
\bibitem [{\citenamefont {Yuan}\ \emph {et~al.}(2016)\citenamefont {Yuan},
  \citenamefont {Zhou}, \citenamefont {He},\ and\ \citenamefont
  {Law}}]{yuan2016ising}%
  \BibitemOpen
  \bibfield  {author} {\bibinfo {author} {\bibfnamefont {N.~F.~Q.}\
  \bibnamefont {Yuan}}, \bibinfo {author} {\bibfnamefont {B.~T.}\ \bibnamefont
  {Zhou}}, \bibinfo {author} {\bibfnamefont {W.-Y.}\ \bibnamefont {He}},\ and\
  \bibinfo {author} {\bibfnamefont {K.~T.}\ \bibnamefont {Law}},\ }\href@noop
  {} {\bibinfo {title} {Ising superconductivity in transition metal
  dichalcogenides}} (\bibinfo {year} {2016}),\ \Eprint
  {https://arxiv.org/abs/1605.01847} {arXiv:1605.01847 [cond-mat.supr-con]}
  \BibitemShut {NoStop}%
\bibitem [{\citenamefont {Liu}\ \emph {et~al.}(2018)\citenamefont {Liu},
  \citenamefont {Wang}, \citenamefont {Zhang}, \citenamefont {Liu},
  \citenamefont {Liu}, \citenamefont {Zhou}, \citenamefont {Wang},
  \citenamefont {Wang}, \citenamefont {Liu}, \citenamefont {Xi}, \citenamefont
  {Tian}, \citenamefont {Liu}, \citenamefont {Feng}, \citenamefont {Xie},\ and\
  \citenamefont {Wang}}]{Liu2018}%
  \BibitemOpen
  \bibfield  {author} {\bibinfo {author} {\bibfnamefont {Y.}~\bibnamefont
  {Liu}}, \bibinfo {author} {\bibfnamefont {Z.}~\bibnamefont {Wang}}, \bibinfo
  {author} {\bibfnamefont {X.}~\bibnamefont {Zhang}}, \bibinfo {author}
  {\bibfnamefont {C.}~\bibnamefont {Liu}}, \bibinfo {author} {\bibfnamefont
  {Y.}~\bibnamefont {Liu}}, \bibinfo {author} {\bibfnamefont {Z.}~\bibnamefont
  {Zhou}}, \bibinfo {author} {\bibfnamefont {J.}~\bibnamefont {Wang}}, \bibinfo
  {author} {\bibfnamefont {Q.}~\bibnamefont {Wang}}, \bibinfo {author}
  {\bibfnamefont {Y.}~\bibnamefont {Liu}}, \bibinfo {author} {\bibfnamefont
  {C.}~\bibnamefont {Xi}}, \bibinfo {author} {\bibfnamefont {M.}~\bibnamefont
  {Tian}}, \bibinfo {author} {\bibfnamefont {H.}~\bibnamefont {Liu}}, \bibinfo
  {author} {\bibfnamefont {J.}~\bibnamefont {Feng}}, \bibinfo {author}
  {\bibfnamefont {X.~C.}\ \bibnamefont {Xie}},\ and\ \bibinfo {author}
  {\bibfnamefont {J.}~\bibnamefont {Wang}},\ }\bibfield  {title} {\bibinfo
  {title} {{Interface-Induced Zeeman-Protected Superconductivity in Ultrathin
  Crystalline Lead Films}},\ }\href {https://doi.org/10.1103/PhysRevX.8.021002}
  {\bibfield  {journal} {\bibinfo  {journal} {Physical Review X}\ }\textbf
  {\bibinfo {volume} {8}},\ \bibinfo {pages} {021002} (\bibinfo {year}
  {2018})}\BibitemShut {NoStop}%
\bibitem [{\citenamefont {woo Cho}\ \emph {et~al.}(2020)\citenamefont {woo
  Cho}, \citenamefont {Lyu}, \citenamefont {Han}, \citenamefont {Ng},
  \citenamefont {Gao}, \citenamefont {Li}, \citenamefont {Huang}, \citenamefont
  {Wang}, \citenamefont {Schmalian},\ and\ \citenamefont {Lortz}}]{Cho2020}%
  \BibitemOpen
  \bibfield  {author} {\bibinfo {author} {\bibfnamefont {C.}~\bibnamefont {woo
  Cho}}, \bibinfo {author} {\bibfnamefont {J.}~\bibnamefont {Lyu}}, \bibinfo
  {author} {\bibfnamefont {T.}~\bibnamefont {Han}}, \bibinfo {author}
  {\bibfnamefont {C.~Y.}\ \bibnamefont {Ng}}, \bibinfo {author} {\bibfnamefont
  {Y.}~\bibnamefont {Gao}}, \bibinfo {author} {\bibfnamefont {G.}~\bibnamefont
  {Li}}, \bibinfo {author} {\bibfnamefont {M.}~\bibnamefont {Huang}}, \bibinfo
  {author} {\bibfnamefont {N.}~\bibnamefont {Wang}}, \bibinfo {author}
  {\bibfnamefont {J.}~\bibnamefont {Schmalian}},\ and\ \bibinfo {author}
  {\bibfnamefont {R.}~\bibnamefont {Lortz}},\ }\href@noop {} {\bibinfo {title}
  {Distinct nodal and nematic superconducting phases in the 2d ising
  superconductor nbse2}} (\bibinfo {year} {2020}),\ \Eprint
  {https://arxiv.org/abs/2003.12467} {arXiv:2003.12467 [cond-mat.supr-con]}
  \BibitemShut {NoStop}%
\bibitem [{\citenamefont {Cho}\ \emph {et~al.}(2022)\citenamefont {Cho},
  \citenamefont {Lyu}, \citenamefont {An}, \citenamefont {Han}, \citenamefont
  {Lo}, \citenamefont {Ng}, \citenamefont {Hu}, \citenamefont {Gao},
  \citenamefont {Li}, \citenamefont {Huang}, \citenamefont {Wang},
  \citenamefont {Schmalian},\ and\ \citenamefont {Lortz}}]{cho22}%
  \BibitemOpen
  \bibfield  {author} {\bibinfo {author} {\bibfnamefont {C.-w.}\ \bibnamefont
  {Cho}}, \bibinfo {author} {\bibfnamefont {J.}~\bibnamefont {Lyu}}, \bibinfo
  {author} {\bibfnamefont {L.}~\bibnamefont {An}}, \bibinfo {author}
  {\bibfnamefont {T.}~\bibnamefont {Han}}, \bibinfo {author} {\bibfnamefont
  {K.~T.}\ \bibnamefont {Lo}}, \bibinfo {author} {\bibfnamefont {C.~Y.}\
  \bibnamefont {Ng}}, \bibinfo {author} {\bibfnamefont {J.}~\bibnamefont {Hu}},
  \bibinfo {author} {\bibfnamefont {Y.}~\bibnamefont {Gao}}, \bibinfo {author}
  {\bibfnamefont {G.}~\bibnamefont {Li}}, \bibinfo {author} {\bibfnamefont
  {M.}~\bibnamefont {Huang}}, \bibinfo {author} {\bibfnamefont
  {N.}~\bibnamefont {Wang}}, \bibinfo {author} {\bibfnamefont {J.}~\bibnamefont
  {Schmalian}},\ and\ \bibinfo {author} {\bibfnamefont {R.}~\bibnamefont
  {Lortz}},\ }\bibfield  {title} {\bibinfo {title} {Nodal and nematic
  superconducting phases in ${\mathrm{nbse}}_{2}$ monolayers from competing
  superconducting channels},\ }\href
  {https://doi.org/10.1103/PhysRevLett.129.087002} {\bibfield  {journal}
  {\bibinfo  {journal} {Phys. Rev. Lett.}\ }\textbf {\bibinfo {volume} {129}},\
  \bibinfo {pages} {087002} (\bibinfo {year} {2022})}\BibitemShut {NoStop}%
\bibitem [{\citenamefont {Frigeri}\ \emph {et~al.}(2006)\citenamefont
  {Frigeri}, \citenamefont {Agterberg}, \citenamefont {Milat},\ and\
  \citenamefont {Sigrist}}]{Frigeri2006}%
  \BibitemOpen
  \bibfield  {author} {\bibinfo {author} {\bibfnamefont {P.~A.}\ \bibnamefont
  {Frigeri}}, \bibinfo {author} {\bibfnamefont {D.~F.}\ \bibnamefont
  {Agterberg}}, \bibinfo {author} {\bibfnamefont {I.}~\bibnamefont {Milat}},\
  and\ \bibinfo {author} {\bibfnamefont {M.}~\bibnamefont {Sigrist}},\
  }\bibfield  {title} {\bibinfo {title} {{Phenomenological theory of the s-wave
  state in superconductors without an inversion center}},\ }\href
  {https://doi.org/10.1140/epjb/e2007-00019-5} {\bibfield  {journal} {\bibinfo
  {journal} {The European Physical Journal B}\ }\textbf {\bibinfo {volume}
  {54}},\ \bibinfo {pages} {435} (\bibinfo {year} {2006})}\BibitemShut
  {NoStop}%
\bibitem [{\citenamefont {Kuzmanovi\ifmmode~\acute{c}\else \'{c}\fi{}}\ \emph
  {et~al.}(2022)\citenamefont {Kuzmanovi\ifmmode~\acute{c}\else \'{c}\fi{}},
  \citenamefont {Dvir}, \citenamefont {LeBoeuf}, \citenamefont
  {Ili\ifmmode~\acute{c}\else \'{c}\fi{}}, \citenamefont {Haim}, \citenamefont
  {M\"ockli}, \citenamefont {Kramer}, \citenamefont {Khodas}, \citenamefont
  {Houzet}, \citenamefont {Meyer}, \citenamefont {Aprili}, \citenamefont
  {Steinberg},\ and\ \citenamefont {Quay}}]{Kuzmanovic2022}%
  \BibitemOpen
  \bibfield  {author} {\bibinfo {author} {\bibfnamefont {M.}~\bibnamefont
  {Kuzmanovi\ifmmode~\acute{c}\else \'{c}\fi{}}}, \bibinfo {author}
  {\bibfnamefont {T.}~\bibnamefont {Dvir}}, \bibinfo {author} {\bibfnamefont
  {D.}~\bibnamefont {LeBoeuf}}, \bibinfo {author} {\bibfnamefont
  {S.}~\bibnamefont {Ili\ifmmode~\acute{c}\else \'{c}\fi{}}}, \bibinfo {author}
  {\bibfnamefont {M.}~\bibnamefont {Haim}}, \bibinfo {author} {\bibfnamefont
  {D.}~\bibnamefont {M\"ockli}}, \bibinfo {author} {\bibfnamefont
  {S.}~\bibnamefont {Kramer}}, \bibinfo {author} {\bibfnamefont
  {M.}~\bibnamefont {Khodas}}, \bibinfo {author} {\bibfnamefont
  {M.}~\bibnamefont {Houzet}}, \bibinfo {author} {\bibfnamefont {J.~S.}\
  \bibnamefont {Meyer}}, \bibinfo {author} {\bibfnamefont {M.}~\bibnamefont
  {Aprili}}, \bibinfo {author} {\bibfnamefont {H.}~\bibnamefont {Steinberg}},\
  and\ \bibinfo {author} {\bibfnamefont {C.~H.~L.}\ \bibnamefont {Quay}},\
  }\bibfield  {title} {\bibinfo {title} {Tunneling spectroscopy of
  few-monolayer ${\mathrm{nbse}}_{2}$ in high magnetic fields: Triplet
  superconductivity and ising protection},\ }\href
  {https://doi.org/10.1103/PhysRevB.106.184514} {\bibfield  {journal} {\bibinfo
   {journal} {Phys. Rev. B}\ }\textbf {\bibinfo {volume} {106}},\ \bibinfo
  {pages} {184514} (\bibinfo {year} {2022})}\BibitemShut {NoStop}%
\bibitem [{\citenamefont {Zhao}\ and\ \citenamefont {Wang}(2013)}]{Zhao2013}%
  \BibitemOpen
  \bibfield  {author} {\bibinfo {author} {\bibfnamefont {Y.~X.}\ \bibnamefont
  {Zhao}}\ and\ \bibinfo {author} {\bibfnamefont {Z.~D.}\ \bibnamefont
  {Wang}},\ }\bibfield  {title} {\bibinfo {title} {Topological classification
  and stability of fermi surfaces},\ }\href
  {https://doi.org/10.1103/PhysRevLett.110.240404} {\bibfield  {journal}
  {\bibinfo  {journal} {Phys. Rev. Lett.}\ }\textbf {\bibinfo {volume} {110}},\
  \bibinfo {pages} {240404} (\bibinfo {year} {2013})}\BibitemShut {NoStop}%
\bibitem [{\citenamefont {Matsuura}\ \emph {et~al.}(2013)\citenamefont
  {Matsuura}, \citenamefont {Chang}, \citenamefont {Schnyder},\ and\
  \citenamefont {Ryu}}]{Matsuura2013}%
  \BibitemOpen
  \bibfield  {author} {\bibinfo {author} {\bibfnamefont {S.}~\bibnamefont
  {Matsuura}}, \bibinfo {author} {\bibfnamefont {P.-Y.}\ \bibnamefont {Chang}},
  \bibinfo {author} {\bibfnamefont {A.~P.}\ \bibnamefont {Schnyder}},\ and\
  \bibinfo {author} {\bibfnamefont {S.}~\bibnamefont {Ryu}},\ }\bibfield
  {title} {\bibinfo {title} {Protected boundary states in gapless topological
  phases},\ }\href {https://doi.org/10.1088/1367-2630/15/6/065001} {\bibfield
  {journal} {\bibinfo  {journal} {New Journal of Physics}\ }\textbf {\bibinfo
  {volume} {15}},\ \bibinfo {pages} {065001} (\bibinfo {year}
  {2013})}\BibitemShut {NoStop}%
\bibitem [{\citenamefont {Galvis}\ \emph {et~al.}(2014)\citenamefont {Galvis},
  \citenamefont {Chirolli}, \citenamefont {Guillam\'on}, \citenamefont
  {Vieira}, \citenamefont {Navarro-Moratalla}, \citenamefont {Coronado},
  \citenamefont {Suderow},\ and\ \citenamefont {Guinea}}]{Galvis2014}%
  \BibitemOpen
  \bibfield  {author} {\bibinfo {author} {\bibfnamefont {J.~A.}\ \bibnamefont
  {Galvis}}, \bibinfo {author} {\bibfnamefont {L.}~\bibnamefont {Chirolli}},
  \bibinfo {author} {\bibfnamefont {I.}~\bibnamefont {Guillam\'on}}, \bibinfo
  {author} {\bibfnamefont {S.}~\bibnamefont {Vieira}}, \bibinfo {author}
  {\bibfnamefont {E.}~\bibnamefont {Navarro-Moratalla}}, \bibinfo {author}
  {\bibfnamefont {E.}~\bibnamefont {Coronado}}, \bibinfo {author}
  {\bibfnamefont {H.}~\bibnamefont {Suderow}},\ and\ \bibinfo {author}
  {\bibfnamefont {F.}~\bibnamefont {Guinea}},\ }\bibfield  {title} {\bibinfo
  {title} {Zero-bias conductance peak in detached flakes of superconducting
  2$h$-${\mathrm{tas}}_{2}$ probed by scanning tunneling spectroscopy},\ }\href
  {https://doi.org/10.1103/PhysRevB.89.224512} {\bibfield  {journal} {\bibinfo
  {journal} {Phys. Rev. B}\ }\textbf {\bibinfo {volume} {89}},\ \bibinfo
  {pages} {224512} (\bibinfo {year} {2014})}\BibitemShut {NoStop}%
\bibitem [{\citenamefont {Nayak}\ \emph {et~al.}(2021)\citenamefont {Nayak},
  \citenamefont {Steinbok}, \citenamefont {Roet}, \citenamefont {Koo},
  \citenamefont {Margalit}, \citenamefont {Feldman}, \citenamefont {Almoalem},
  \citenamefont {Kanigel}, \citenamefont {Fiete}, \citenamefont {Yan},
  \citenamefont {Oreg}, \citenamefont {Avraham},\ and\ \citenamefont
  {Beidenkopf}}]{Nayak2021}%
  \BibitemOpen
  \bibfield  {author} {\bibinfo {author} {\bibfnamefont {A.~K.}\ \bibnamefont
  {Nayak}}, \bibinfo {author} {\bibfnamefont {A.}~\bibnamefont {Steinbok}},
  \bibinfo {author} {\bibfnamefont {Y.}~\bibnamefont {Roet}}, \bibinfo {author}
  {\bibfnamefont {J.}~\bibnamefont {Koo}}, \bibinfo {author} {\bibfnamefont
  {G.}~\bibnamefont {Margalit}}, \bibinfo {author} {\bibfnamefont
  {I.}~\bibnamefont {Feldman}}, \bibinfo {author} {\bibfnamefont
  {A.}~\bibnamefont {Almoalem}}, \bibinfo {author} {\bibfnamefont
  {A.}~\bibnamefont {Kanigel}}, \bibinfo {author} {\bibfnamefont {G.~A.}\
  \bibnamefont {Fiete}}, \bibinfo {author} {\bibfnamefont {B.}~\bibnamefont
  {Yan}}, \bibinfo {author} {\bibfnamefont {Y.}~\bibnamefont {Oreg}}, \bibinfo
  {author} {\bibfnamefont {N.}~\bibnamefont {Avraham}},\ and\ \bibinfo {author}
  {\bibfnamefont {H.}~\bibnamefont {Beidenkopf}},\ }\bibfield  {title}
  {\bibinfo {title} {Evidence of topological boundary modes with topological
  nodal-point superconductivity},\ }\href
  {https://doi.org/10.1038/s41567-021-01376-z} {\bibfield  {journal} {\bibinfo
  {journal} {Nature Physics}\ }\textbf {\bibinfo {volume} {17}},\ \bibinfo
  {pages} {1413} (\bibinfo {year} {2021})}\BibitemShut {NoStop}%
\bibitem [{\citenamefont {Seshadri}\ \emph {et~al.}(2022)\citenamefont
  {Seshadri}, \citenamefont {Khodas},\ and\ \citenamefont {Meidan}}]{sesh22}%
  \BibitemOpen
  \bibfield  {author} {\bibinfo {author} {\bibfnamefont {R.}~\bibnamefont
  {Seshadri}}, \bibinfo {author} {\bibfnamefont {M.}~\bibnamefont {Khodas}},\
  and\ \bibinfo {author} {\bibfnamefont {D.}~\bibnamefont {Meidan}},\
  }\bibfield  {title} {\bibinfo {title} {{Josephson junctions of topological
  nodal superconductors}},\ }\href
  {https://doi.org/10.21468/SciPostPhys.12.6.197} {\bibfield  {journal}
  {\bibinfo  {journal} {SciPost Phys.}\ }\textbf {\bibinfo {volume} {12}},\
  \bibinfo {pages} {197} (\bibinfo {year} {2022})}\BibitemShut {NoStop}%
\bibitem [{\citenamefont {Shaffer}\ \emph {et~al.}(2020)\citenamefont
  {Shaffer}, \citenamefont {Kang}, \citenamefont {Burnell},\ and\ \citenamefont
  {Fernandes}}]{fernandes20}%
  \BibitemOpen
  \bibfield  {author} {\bibinfo {author} {\bibfnamefont {D.}~\bibnamefont
  {Shaffer}}, \bibinfo {author} {\bibfnamefont {J.}~\bibnamefont {Kang}},
  \bibinfo {author} {\bibfnamefont {F.~J.}\ \bibnamefont {Burnell}},\ and\
  \bibinfo {author} {\bibfnamefont {R.~M.}\ \bibnamefont {Fernandes}},\
  }\bibfield  {title} {\bibinfo {title} {Crystalline nodal topological
  superconductivity and bogolyubov fermi surfaces in monolayer
  ${\mathrm{nbse}}_{2}$},\ }\href {https://doi.org/10.1103/PhysRevB.101.224503}
  {\bibfield  {journal} {\bibinfo  {journal} {Phys. Rev. B}\ }\textbf {\bibinfo
  {volume} {101}},\ \bibinfo {pages} {224503} (\bibinfo {year}
  {2020})}\BibitemShut {NoStop}%
\bibitem [{\citenamefont {Altland}\ and\ \citenamefont
  {Zirnbauer}(1997)}]{alt97}%
  \BibitemOpen
  \bibfield  {author} {\bibinfo {author} {\bibfnamefont {A.}~\bibnamefont
  {Altland}}\ and\ \bibinfo {author} {\bibfnamefont {M.~R.}\ \bibnamefont
  {Zirnbauer}},\ }\bibfield  {title} {\bibinfo {title} {Nonstandard symmetry
  classes in mesoscopic normal-superconducting hybrid structures},\ }\href
  {https://doi.org/10.1103/PhysRevB.55.1142} {\bibfield  {journal} {\bibinfo
  {journal} {Phys. Rev. B}\ }\textbf {\bibinfo {volume} {55}},\ \bibinfo
  {pages} {1142} (\bibinfo {year} {1997})}\BibitemShut {NoStop}%
\bibitem [{\citenamefont {Haim}\ \emph {et~al.}(2022)\citenamefont {Haim},
  \citenamefont {Levchenko},\ and\ \citenamefont {Khodas}}]{Haim2022}%
  \BibitemOpen
  \bibfield  {author} {\bibinfo {author} {\bibfnamefont {M.}~\bibnamefont
  {Haim}}, \bibinfo {author} {\bibfnamefont {A.}~\bibnamefont {Levchenko}},\
  and\ \bibinfo {author} {\bibfnamefont {M.}~\bibnamefont {Khodas}},\
  }\bibfield  {title} {\bibinfo {title} {Mechanisms of in-plane magnetic
  anisotropy in superconducting ${\mathrm{nbse}}_{2}$},\ }\href
  {https://doi.org/10.1103/PhysRevB.105.024515} {\bibfield  {journal} {\bibinfo
   {journal} {Phys. Rev. B}\ }\textbf {\bibinfo {volume} {105}},\ \bibinfo
  {pages} {024515} (\bibinfo {year} {2022})}\BibitemShut {NoStop}%
\bibitem [{\citenamefont {Smidman}\ \emph
  {et~al.}(2017{\natexlab{b}})\citenamefont {Smidman}, \citenamefont {Salamon},
  \citenamefont {Yuan},\ and\ \citenamefont
  {Agterberg}}]{smidman2017superconductivity}%
  \BibitemOpen
  \bibfield  {author} {\bibinfo {author} {\bibfnamefont {M.}~\bibnamefont
  {Smidman}}, \bibinfo {author} {\bibfnamefont {M.}~\bibnamefont {Salamon}},
  \bibinfo {author} {\bibfnamefont {H.}~\bibnamefont {Yuan}},\ and\ \bibinfo
  {author} {\bibfnamefont {D.}~\bibnamefont {Agterberg}},\ }\bibfield  {title}
  {\bibinfo {title} {Superconductivity and spin--orbit coupling in
  non-centrosymmetric materials: a review},\ }\href@noop {} {\bibfield
  {journal} {\bibinfo  {journal} {Reports on Progress in Physics}\ }\textbf
  {\bibinfo {volume} {80}},\ \bibinfo {pages} {036501} (\bibinfo {year}
  {2017}{\natexlab{b}})}\BibitemShut {NoStop}%
\bibitem [{\citenamefont {Hughes}\ \emph {et~al.}(2011)\citenamefont {Hughes},
  \citenamefont {Prodan},\ and\ \citenamefont {Bernevig}}]{Hughes2011}%
  \BibitemOpen
  \bibfield  {author} {\bibinfo {author} {\bibfnamefont {T.~L.}\ \bibnamefont
  {Hughes}}, \bibinfo {author} {\bibfnamefont {E.}~\bibnamefont {Prodan}},\
  and\ \bibinfo {author} {\bibfnamefont {B.~A.}\ \bibnamefont {Bernevig}},\
  }\bibfield  {title} {\bibinfo {title} {Inversion-symmetric topological
  insulators},\ }\href {https://doi.org/10.1103/PhysRevB.83.245132} {\bibfield
  {journal} {\bibinfo  {journal} {Phys. Rev. B}\ }\textbf {\bibinfo {volume}
  {83}},\ \bibinfo {pages} {245132} (\bibinfo {year} {2011})}\BibitemShut
  {NoStop}%
\bibitem [{\citenamefont {Chiu}\ \emph {et~al.}(2013)\citenamefont {Chiu},
  \citenamefont {Yao},\ and\ \citenamefont {Ryu}}]{Chiu2013}%
  \BibitemOpen
  \bibfield  {author} {\bibinfo {author} {\bibfnamefont {C.-K.}\ \bibnamefont
  {Chiu}}, \bibinfo {author} {\bibfnamefont {H.}~\bibnamefont {Yao}},\ and\
  \bibinfo {author} {\bibfnamefont {S.}~\bibnamefont {Ryu}},\ }\bibfield
  {title} {\bibinfo {title} {Classification of topological insulators and
  superconductors in the presence of reflection symmetry},\ }\href
  {https://doi.org/10.1103/PhysRevB.88.075142} {\bibfield  {journal} {\bibinfo
  {journal} {Phys. Rev. B}\ }\textbf {\bibinfo {volume} {88}},\ \bibinfo
  {pages} {075142} (\bibinfo {year} {2013})}\BibitemShut {NoStop}%
\bibitem [{\citenamefont {Laubscher}\ and\ \citenamefont
  {Klinovaja}(2021)}]{klin21}%
  \BibitemOpen
  \bibfield  {author} {\bibinfo {author} {\bibfnamefont {K.}~\bibnamefont
  {Laubscher}}\ and\ \bibinfo {author} {\bibfnamefont {J.}~\bibnamefont
  {Klinovaja}},\ }\bibfield  {title} {\bibinfo {title} {Majorana bound states
  in semiconducting nanostructures},\ }\href
  {https://doi.org/10.1063/5.0055997} {\bibfield  {journal} {\bibinfo
  {journal} {Journal of Applied Physics}\ }\textbf {\bibinfo {volume} {130}},\
  \bibinfo {pages} {081101} (\bibinfo {year} {2021})},\ \Eprint
  {https://arxiv.org/abs/https://doi.org/10.1063/5.0055997}
  {https://doi.org/10.1063/5.0055997} \BibitemShut {NoStop}%
\end{thebibliography}%
\end{document}